\newif\ifconfver
\confverfalse      

\newif\ifcutshort      
\cutshorttrue

\newif\ifcutshortlvltwo  
\cutshortlvltwofalse

\ifconfver
\documentclass[10pt,twocolumn,twoside]{IEEEtran}
\else
\documentclass[12pt,draftclsnofoot,onecolumn]{IEEEtran}
\fi

\usepackage{array}
\usepackage{cite}
\usepackage{calc}
\usepackage{caption}
\usepackage{graphicx}
\usepackage{psfrag}
\usepackage{stfloats}
\usepackage{url}
\usepackage{booktabs,siunitx}
\usepackage{subfigure,relsize}
\usepackage{longtable}
\usepackage{amsmath,amsfonts,amssymb,amsbsy,bm,paralist,theorem,ifthen,color}
\usepackage{mathtools}
\usepackage{algorithmic,algorithm}
\usepackage[dvips]{epsfig}
\usepackage[dvips]{graphicx}
\usepackage{caption}
\usepackage[font={small}]{caption}

\definecolor{orange}{RGB}{255,107,0}

\definecolor{green}{RGB}{0,180,80}

\newcommand\Yc{\ensuremath{\mathcal{Y}}}

\newcommand\Sc{\ensuremath{\mathcal{S}}}

\newcommand\Lc{\ensuremath{{\mathcal{L}}}}

\newcommand\Kc{\ensuremath{{\mathcal{K}}}}

\newcommand\xb{\ensuremath{{\bm x}}}

\newcommand\yb{\ensuremath{{\bm y}}}

\newcommand\ssb{\ensuremath{{\bm s}}}
\renewcommand\sb{\ensuremath{{\bf s}}}
\newcommand\ub{\ensuremath{{\bm u}}}

\newcommand\ab{\ensuremath{{\bm a}}}

\newcommand\bb{\ensuremath{{\bm b}}}

\newcommand\db{\ensuremath{{\bm d}}}

\newcommand\eb{\ensuremath{{\bm e}}}

\newcommand\Ib{\ensuremath{{\bm I}}}

\newcommand\pb{\ensuremath{{\bm p}}}

\newcommand\tb{\ensuremath{{\bm t}}}

\newcommand\zb{\ensuremath{{\bm z}}}

\newcommand\nub{\ensuremath{{\bm \nu}}}

\newcommand\zerob{\ensuremath{{\bm 0}}}

\newcommand\oneb{\ensuremath{{\bf 1}}}

\newtheorem{prop}{Proposition}
\newtheorem{remark}{Remark}
\newtheorem{property}{Property}

\graphicspath{{figure/}}
\begin{document}
	
\bibliographystyle{IEEEtran}

\title{UAV Positioning and Power Control for Two-Way Wireless Relaying}

\ifconfver \else {\linespread{1.1} \rm \fi
	
\author{\IEEEauthorblockN{Lei Li\IEEEauthorrefmark{1},
Tsung-Hui Chang\IEEEauthorrefmark{2}, and
Shu Cai\IEEEauthorrefmark{3},}

\thanks{ T.-H. Chang is with the School of Science and Engineering, The Chinese University of Hong Kong, Shenzhen, China, and with the Shenzhen Research Institute of Big Data (email: tsunghui.chang@ieee.org). 

L. Li is with the Virginia Polytechnic Institute and State University. He was with the School of Science and Engineering, The Chinese University of Hong Kong, Shenzhen, China. (email: lei.ap@outlook.com).

S. Cai is with the Nanjing University of Posts and Telecommunications, Nanjing, China (email: caishu@njupt.edu.cn).}
}

\maketitle
\begin{abstract}
	This paper considers an unmanned-aerial-vehicle-enabled (UAV-enabled) wireless network where a relay UAV is used for two-way communications between a ground base station (BS) and a set of distant user equipment (UE). The UAV adopts the amplify-and-forward strategy for two-way relaying over orthogonal frequency bands.
	The UAV positioning and the transmission powers of all nodes are jointly designed to maximize the sum rate of both uplink and downlink subject to transmission power constraints and the signal-to-noise ratio constraint on the UAV control channel. The formulated joint positioning and power control (JPPC) problem has an intricate expression of the sum rate due to two-way transmissions and is difficult to solve in general. We propose a novel concave surrogate function for the sum rate and employ the successive convex approximation (SCA) technique for obtaining a high-quality approximate solution. We show that the proposed surrogate function has a small curvature and enables a fast convergence of SCA. Furthermore, we develop a computationally efficient JPPC algorithm by applying the FISTA-type accelerated gradient projection (AGP) algorithm to solve the SCA problem as well as one of the projection subproblem, resulting in a double-loop AGP method. Simulation results show that the proposed JPPC algorithms are not only computationally efficient but also  greatly outperform the heuristic approaches.
\end{abstract}
\ifconfver \else
\newpage
\fi
\ifconfver \else \IEEEpeerreviewmaketitle} \fi


\section{Introduction}
	Recently, deploying unmanned aerial vehicles (UAVs) in wireless communication networks for coverage and throughput enhancement has attracted significant attention from both the industry and academia \cite{zeng2016wireless,Tutorial,qualcomm}. 
	The swift mobility of UAV enables fast deployment and establishment of communications in emergency situations such as for rescue after hurricane or earthquake.
	The lower cost of UAV than the traditional communication infrastructure also makes UAV a cost-effective option for the network coverage and throughput enhancement in coverage-limited zones like the rural or mountainous areas. 
	Besides, UAVs in general have
	better air-to-ground (A2G) channels due to a high probability of line
	of sight (LOS) link with ground users \cite{sun2015dual}.
	Therefore, the UAV has been considered for being an aerial base station (BS) \cite{Y.Z_TWC17,arXiv1801,H.He_CL18,J.Lyu_Spiral,Q.WU_TWC18,C.Zhang_JSAC17}, wireless relay \cite{WirelessDay17,TAES11,Y.Z_TWC17_AO,arXiv1801_L,USC,ICC16}, and for networking \cite{ZhuHan,A.C_TMC18} as well as for data collection and dissemination in wireless sensor networks \cite{Dong2014uav,jiang2012optimization,Adbulla2014optimal,GongChang018,ShenChang018}.
	Several industrial projects that leverage the UAV for enhanced wireless communications, like the Facebook’s laser drone test \cite{facebook} and Qualcomm’s drone communication plan \cite{qualcomm}, are also proposed.

\vspace{-1em}
\subsection{Related Works}
	There are still many technical challenges to overcome in order to harvest the benefits of UAV-enabled wireless communications \cite{Tutorial}. Specifically, the air-to-ground (A2G) channel is different from the existing ground-to-ground channel, and is highly dependent on the position of UAV. 
	In addition, due to limited battery energy, joint positioning/flying trajectory design and transmission power control are critical to achieve high spectral efficiency and energy efficiency in UAV-enabled communication systems. 
	For example, reference \cite{Y.Z_TWC17} derived a fix-wing UAV propulsion energy consumption model and studied the joint UAV trajectory and transmission power control problem for maximizing the system energy efficiency.
	By deploying the UAV as an aerial BS, reference \cite{arXiv1801} studied the trajectory and power control problem for maximizing the minimum downlink rate of ground users over orthogonal channels.
	By assuming that the aerial BS has multiple antennas, reference \cite{H.He_CL18} considered joint optimization of the UAV flying altitude and beamwidth for throughput maximization in multicast, broadcast and uplink scenarios, respectively.   
	Reference \cite{J.Lyu_Spiral} considered the placement of a minimum number UAV-mounted BSs for providing required quality of service for the ground users, while \cite{Q.WU_TWC18} studied the joint scheduling, flying trajectory and power control of multiple UAV-mounted BSs for maximizing the minimum rate of served ground users.
	Unlike \cite{J.Lyu_Spiral,Q.WU_TWC18}, by modeling the positions of the UAVs as a 3-dimensional Poisson point process,    
	the work of \cite{C.Zhang_JSAC17} considered the spectrum sharing problem between the cellular network and drone small cells, and investigated the deployment density of UAVs to maximize the outage-constrained throughput.
	While most of the aforementioned works have assumed deterministic LOS links, the work \cite{M.M_global15} has studied the optimal flying altitude of a UAV for coverage maximization under a probabilistic LOS channel model \cite{A.A_WCL14}.
	
	When the UAV is deployed as a wireless relay, the position and flying trajectory design are also of great importance  \cite{WirelessDay17}.
	The work \cite{TAES11} considered an uplink relaying system and optimized the flying heading of the UAV for maximizing an ergodic transmission rate. 
	In \cite{Y.Z_TWC17_AO}, a decode-and-forward relay system is considered, and the UAV flying trajectory and transmission power are jointly optimized for maximizing the throughput between the ground BS and user equipment (UE).
	In \cite{USC}, the authors considered the UAV positioning problem in a relay system by incorporating the local topological information, where the UAV is aimed to be deployed in a position that can enjoy LOS links.
	The work \cite{arXiv1801_L} considered an uplink multi-UAV relaying system under the LOS channels with random phase. The UAV positions and UE transmission powers are jointly optimized to maximize the minimum ergodic throughput of ground UEs. 
	Reference \cite{ICC16} considered the use of a relay UAV for communicating with another observation UAV and studied the optimal positioning of the relay UAV for throughput maximization.
	The works \cite{ZhuHan} and  \cite{A.C_TMC18} considered the deployment of multiple relay UAVs to form an ad-hoc network and achieve long distance communications, respectively.

\vspace{-1em}
\subsection{Contributions}
\begin{figure}[t!]
		\centering
		\includegraphics[width=0.55\linewidth]{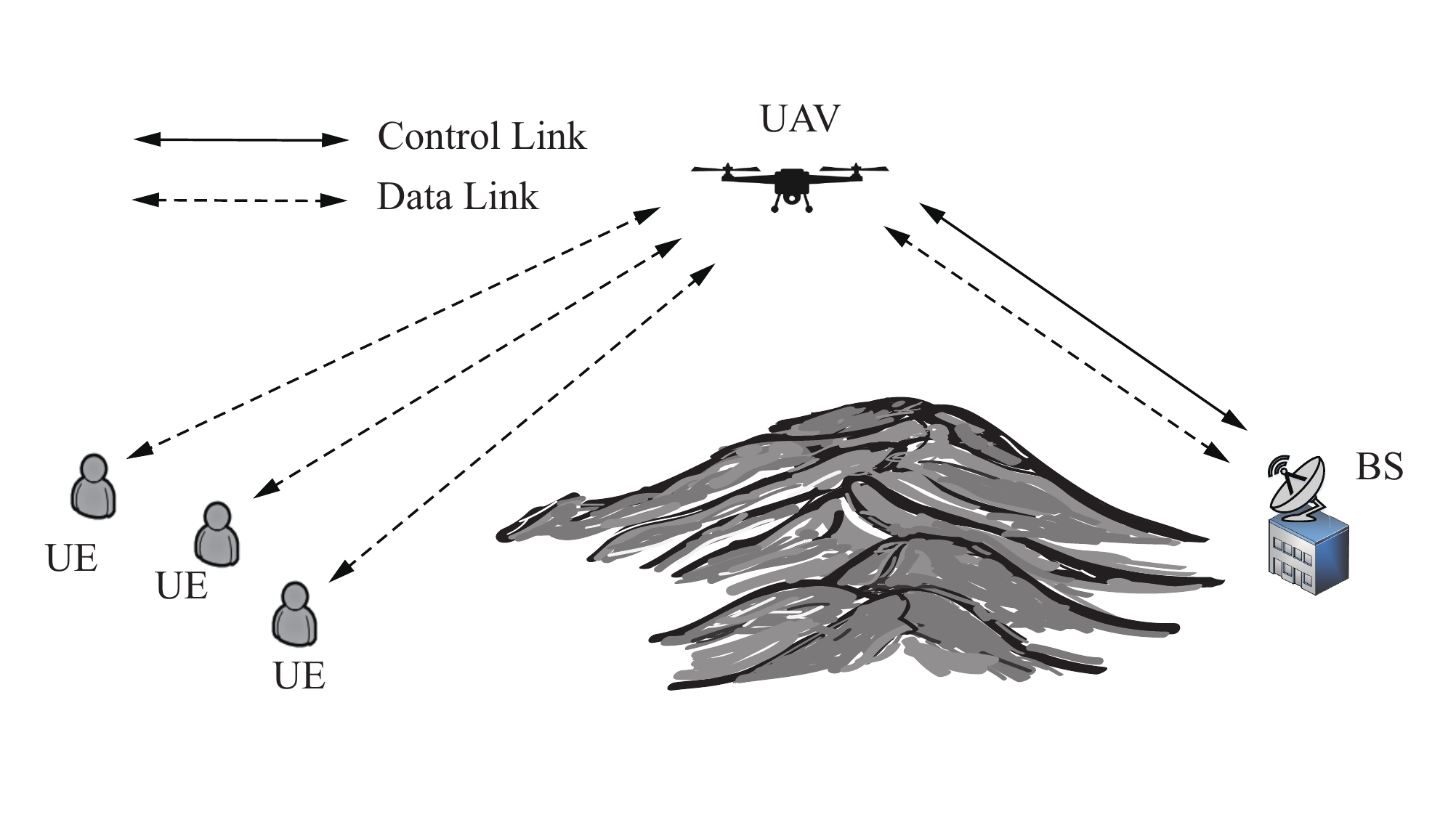}
		\vspace{-0.5cm}
		\centering	\caption{UAV-enabled two-way relay communications}
		\label{fig:fig1}
		\vspace{-0.6cm}
	\end{figure}

	In this paper, we consider a wireless relay network where the UAV is used to extend the service of a BS for a set of distant ground UEs, as shown in Fig. \ref{fig:fig1}. Different from the aforementioned works where either uplink or downlink transmission is considered, we consider the two-way communications between the BS and ground UEs.
	Besides, unlike  \cite{TAES11,arXiv1801} which consider only one-hop communication between the UEs and UAV, we consider 
	the two-hop communications where the relay UAV amplifies and forwards (AF) the signals from one side to the other.
	
	We assume the LOS channels and aim to optimize the UAV position and transmission powers of the BS, UEs and the UAV jointly, for maximizing the sum rate of the two-way communication links.
	Except for the maximum transmission power constraints, we also consider the quality of service (QoS) constraint on the control link between the BS and the relay UAV. In practice, the control link is used for control and command signaling between the relay UAV and the BS, and is essential to the UAV motion control.
	The formulated joint UAV positioning and power control (JPPC) problem has a complicated non-concave sum rate function and is difficult to solve in general. The main contributions are summarized as below.
	
	\begin{enumerate}		
		\item We first consider a simple scenario with only one UE \cite{WirelessDay17}, and present a semi-analytical solution to the JPPC problem. It is shown that the optimal position of the relay UAV, when projected onto the x-y plane,  must lie on the line segment between the BS and UE.
		
		\item For the general case with multiple UEs, we employ the successive convex approximation (SCA) technique \cite{SCA_1}. In SCA, one solves a surrogate convex optimization problem iteratively by replacing the non-concave objective by a concave surrogate function.
		Interestingly, according to \cite{Razaviyayn_parallelBCD14}, the curvature of the surrogate function has a direct impact on the convergence behavior of the SCA iterations. By carefully exploiting the function structure, we propose a concave surrogate function for the SCA algorithm. Moreover, we show that the proposed surrogate function has a smaller curvature than the one that is obtained by following the recent work \cite{ShenChang018}, and can lead to a fast convergence of the SCA iterations. 
		
		\item The SCA algorithm requires one to globally solve the convex surrogate problem in every iteration. Since the convex surrogate problem does not admit closed-from solutions, it requires one to employ another powerful optimization method in order to solve the surrogate problem, which may not be efficient especially when the number of UEs is large.
		To improve the computation efficiency, we adopt a recently proposed algorithm by
		\cite{Shao:2018uj} which combines the SCA iteration with the FISTA-type accelerated gradient projection (AGP) algorithm \cite{BeckFISTA2009}.
		By applying the algorithm in \cite{Shao:2018uj} to our JPPC problem, one of the step involves projection onto a set of quadratic constraints. We further employ the AGP method to solve the Lagrange dual problem of the projection step. Thus, the proposed algorithm for the JPPC problem involves double loops of AGP iterations. 
		
		\item Simulation results are presented to show that the SCA algorithm using the proposed surrogate function exhibits a significantly faster convergence behavior than that using \cite{ShenChang018}. Besides, the double-loop AGP algorithm can further reduce the computation time by more than an order of magnitude.
		Simulation results also reveal that fact that the optimal positioning of the relay UAV is not trivial since the optimized solution can greatly outperform simple strategies that deploy the relay UAV on top of the BS or in a geometric center of the network.		
	\end{enumerate}

	The remainder of the paper is organized as follows. 
	Section {\ref{sec:sytem_model}} presents the two-way relay system model and formulates the JPPC problem. 
	The scenario with only one UE is studied in Section {\ref{sec:sigUE}}. 
	In Section {\ref{sec:mulUE}}, the proposed SCA algorithm and double-loop AGP algorithm are presented.
	The simulation results are given in Section {\ref{sec:simu}}. Finally, conclusions are drawn in Section {\ref{sec:conclusion}}.

\section{System Model and Problem Formulation} \label{sec:sytem_model}
\subsection{System Model} \label{subsec:sytem_model}
	As illustrated in Fig. \ref{fig:fig1}, we consider a UAV-enabled wireless two-way relaying communication network constituted by $K$ UEs, one UAV and one BS.
	All the nodes are equipped with single antenna. 
	It is assumed that there is no direct communication link between the UEs and the BS, and the UAV plays a role relaying the uplink signals from the UEs to the BS as well as relaying the downlink signals from the BS to the UEs. 
	So the UAV extends the service coverage of the BS, and its flying and communication are controlled by the BS. 
	Without lose of generality, we assume that all the UEs and the BS are located at the same ground plane. Denote by $\ub_k = (x_k, y_k, 0)^T \in \mathbb{R}^3$ a three dimension (3D) location of the $k$th UE and by $\bb = (x_b, y_b, 0)^T \in \mathbb{R}^3 $ the 3D location of the BS. The UAV flies in the sky with a fixed altitude $h$ (meters) and its 3D location is denoted by 
	$\xb_r = (x_r, y_r, h)^T \in \mathbb{R}^3$.
	
	We assume that the frequency division duplex (FDD) is used for uplink and downlink communications. 
	The UAV works as a two-way relay which amplifies and forwards the uplink and downlink signals to the BS and UEs, respectively.
	Besides, the frequency division multiplxing (FDM) is used so that the communication links of different UEs are orthogonal to each other and have no cross-link interference.
	For the uplink transmission, i.e., the UE$\rightarrow$UAV$\rightarrow$BS link, we denote $p_{u,k} \ge 0$ as the transmission power of each UE $k$, where $k \in \mathcal{K} \triangleq \left\lbrace 1, \cdots, K\right\rbrace$. The transmission power allocated by the UAV for relaying the uplink signals from UE $k$ to the BS is denoted by $p_{r,k}^{\rm U} \ge 0, k \in \mathcal{K} $. For the downlink transmission, i.e., the BS$\rightarrow$UAV$\rightarrow$UE link, the transmission power of the BS for UE $k$ is $p_{b,k} \ge 0, k \in \mathcal{K}$. The downlink relaying power of the UAV for UE $k$ is denoted as $p_{r,k}^{\rm D} \ge 0, k \in \mathcal{K}$. Since the air-to-ground (A2G) channel between the UAV and BS and that between the UAV and UEs usually consist of a strong line-of-sight (LOS) link \cite{arXiv1801_L}, we adopt this model throughout this paper.
	
	{\bf Uplink signal model:} Denote $s_k^{\rm U} \sim \mathcal{N}(0,1)$ as the Gaussian information signal sent by UE $k$.
	In the first time slot of the AF transmission,  the signals received by the UAV are given by
	\begin{equation} \label{rx:ul:relay}
	y_{r,k}^{\rm U} = \sqrt{\frac{ \beta p_{u,k}}{d_{kr}^{2}}}s_k^{\rm U}+ w_r,~k\in  \Kc,
	\end{equation} 
	where $\beta$ is the reference channel gain at the distance $1$ meter from the UE, $d_{kr} \triangleq 
	\|\xb_r - \ub_k\|$ is the Euclidean distance between UE $k$ and the UAV, and $w_r\sim \mathcal{N}\left(0,\sigma^2 \right)$ is the additive noise with zero mean and variance $\sigma^2$.
	In the second time slot, the UAV amplifies the received signal $y_{r,k}^{\rm U}$ and transmits it to the BS. 
	In particular, by assuming that the channel state information (CSI) is available at the UAV, the UAV can amplify the signal $y_{r,k}^{\rm U}$ with 
	a gain $\sqrt{p_{r,k}^{\rm U}g_{r,k}^{\rm U}}$ where 
	\begin{align}\label{gain u}
	g_{r,k}^{\rm U} = \frac{1}{{p_{u,k} \beta d_{kr}^{-2} + \sigma^2}}
	\end{align} is the inverse of the signal power of $y_{r,k}^{\rm U}$, and $p_{r,k}^{\rm U} >0$ is the uplink transmission power of the UAV. Thus the received signal at the BS for UE $k$ is given by
	\begin{equation} \label{rx:ul:BS}
	\begin{split}
	y_{b,k}^{\rm U} & =  \sqrt{ \frac{\beta }{d_{rb}^{2}} } \sqrt{  p_{r,k}^{\rm U} g_{r,k}^{\rm U}}~ y_{r,k}^{\rm U} + w_b\\
	& =  \sqrt{ \frac{\beta }{d_{rb}^{2}} } \sqrt{  p_{r,k}^{\rm U} g_{r,k}^{\rm U}} \left( \sqrt{ \frac{\beta p_{u,k}} {d_{kr}^{2}}}s_k^{\rm U}+ w_r \right) + w_b \\
	& = \sqrt{ \frac{\beta^2 p_{r,k}^{\rm U} p_{u,k}}{d_{rb}^{2} d_{kr}^{2}}  g_{r,k}^{\rm U}} s_k^{\rm U} + \sqrt{ \frac{\beta p_{r,k}^{\rm U} }{d_{rb}^{2}}  g_{r,k}^{\rm U}} w_r + w_b,	
	\end{split}
	\end{equation}
	where $d_{rb} \triangleq 
	\|\xb_r - \bb\|$ is the distance between the UAV and the BS and $w_b\sim \mathcal{N}\left(0,\sigma^2 \right)$ is the additive noise at the BS. 
	By \eqref{rx:ul:BS}, the uplink signal-to-noise ratio (SNR) for the $k$th UE can be expressed as
	\begin{align}
	{\rm SNR}_k^{\rm U} & = \frac{ \beta^2 g_{r,k}^{\rm U} p_{r,k}^{\rm U}  p_{u,k} d_{rb}^{-2} d_{kr}^{-2}}{\beta g_{r,k}^{\rm U} p_{r,k}^{\rm U} d_{rb}^{-2}\sigma^2 + \sigma^2} \notag  \\
	& = \frac{\beta^2 p_{r,k}^{\rm U}  p_{u,k}  d_{kr}^{-2} d_{rb}^{-2}}{\beta p_{r,k}^{\rm U} d_{rb}^{-2}\sigma^2 + \beta p_{u,k} d_{kr}^{-2}\sigma^2 + (\sigma^2)^2} \notag \\
	& =  \frac{  p_{r,k}^{\rm U}  p_{u,k} d_{kr}^{-2} d_{rb}^{-2} \xi  }{p_{r,k}^{\rm U} d_{rb}^{-2} +  p_{u,k} d_{kr}^{-2} + \xi^{-1}} ,  \label{snr:ul:BS}
	\end{align}
	where \eqref{gain u} is applied to obtain the second equality and $\xi \triangleq \frac{\beta}{\sigma^2}$ is defined in the last equality.
	
	{\bf Downlink signal model:} 
	In the downlink transmission, given the information signal $s_k^{\rm D} \sim \mathcal{N}(0,1)$ for UE $k$ sent from the BS in the first time slot, 
	the received signal at the UAV is given by 
	\begin{equation} \label{rx:dl:relay}
	y_{r,k}^{\rm D} = \sqrt{\frac{ \beta p_{b,k}}{d_{rb}^{2}}}s_k^{\rm D}+ w_r.
	\end{equation} 
	In the second time slot, the UAV amplifies $y_{r,k}^{\rm D}$ by the gain $\sqrt{p_{r,k}^{\rm D}g_{r,k}^{\rm D}}$, where
	\begin{align}
	g_{r,k}^{\rm D} = \frac{1}{{\beta  p_{b,k} d_{rb}^{-2} + \sigma^2}} ,
	\end{align} and forwards it to UE $k$. 
	The received signal at UE $k$ is given by
	\begin{equation} \label{rx:dl:UE}
	\begin{split}
	y_{b,k}^{\rm U} & =  \sqrt{ \frac{\beta  }{d_{kr}^{2}} } \sqrt{ p_{r,k}^{\rm D} g_{r,k}^{\rm D}}~ y_{r,k}^{\rm D} + w_k\\
	& = \sqrt{ \frac{\beta^2 p_{r,k}^{\rm D} p_{b,k}}{d_{kr}^{2} d_{rb}^{2}}  g_{r,k}^{\rm D}} s_k^{\rm D} + \sqrt{ \frac{\beta p_{r,k}^{\rm D} }{d_{kr}^{2}}  g_{r,k}^{\rm D} }w_r + w_k,	
	\end{split}
	\end{equation}
	where  $w_k\sim \mathcal{N}\left(0,\sigma^2 \right)$ is the additive noise at UE $k$. 
	By \eqref{rx:dl:UE}, the downlink SNR for the $k$th UE is thus given by 
	\begin{equation} \label{snr:dl:UE}
	{\rm SNR}_k^{\rm D} = \frac{  p_{r,k}^{\rm D}  p_{b,k} d_{rb}^{-2} d_{kr}^{-2} \xi }{ p_{r,k}^{\rm D} d_{kr}^{-2} +   p_{b,k} d_{rb}^{-2} + \xi^{-1}}.
	\end{equation}
	
	Denote by 
	${\pb_{r}^{\rm U}} \triangleq (p_{r,1}^{\rm U},\cdots,p_{r,K}^{\rm U})^T$, ${\pb_{r}^{\rm D}} \triangleq (p_{r,1}^{\rm D},\cdots,p_{r,K}^{\rm D})^T$, ${\pb_{b}} \triangleq (p_{b,1},\cdots,p_{b,K})^T$ and ${\pb_{u}} \triangleq (p_{u,1},\cdots,p_{u,K})^T$ the vectors that collect the transmission powers of the UAV, BS and the UEs, respectively. 
	Based on the uplink and downlink SNR expressions in (\ref{snr:ul:BS}) and (\ref{snr:dl:UE}), the sum rate of the network is 
	\begin{equation} \label{sRate}
	\begin{split}
	R_s(\xb_r, \pb_b,\pb_r^{\rm U},\pb_r^{\rm D},\pb_u) & = \sum_{k=1}^K ( R_k^{\rm U}(\xb_r, \pb_r^{\rm U},\pb_u) + R_k^{\rm D}(\xb_r, \pb_b,\pb_r^{\rm D}))\\
	& = \sum_{k=1}^K \frac{W}{2}\left(  \text{log}(1+ {\rm SNR}_k^{\rm U}) +  \text{log}(1+ {\rm SNR}_k^{\rm D}) \right) ,
	\end{split}
	\end{equation}
	where $W$ is the frequency bandwidth allocated for each UE, and $R_k^{\rm U}(\xb_r, \pb_r^{\rm U},\pb_u)=\frac{W}{2} \text{log}(1+ {\rm SNR}_k^{\rm U})$ and $R_k^{\rm D}(\xb_r, \pb_b, \pb_r^{\rm D})=\frac{W}{2} \text{log}(1+ {\rm SNR}_k^{\rm D})$ are respectively the uplink and downlink transmission rates of each UE $k$. 
	As the AF relay transmission requires two time slots, the rate is divided by 2 in \eqref{sRate}.
	
	{\bf Control link:} 
	Besides the data transmission, signaling on the control link between the UAV and the BS requires stringent communication quality. 
	Let us denote $p_{c}$ as the trasmisssion power for the control signaling between the BS and the UAV. Then, the received SNR for the control link is 
	\begin{equation} \label{snr:ctrl:dl}
	{\rm SNR}_{c} = \frac{\beta p_{c}}{d_{rb}^2 \sigma^2} =  \frac{\xi p_{c}}{d_{rb}^2}.
	\end{equation}
	Note that the control link is symmetric between the BS and the UAV under the LOS channel model. Thus, both the UAV and BS use the same power $p_{c}$ for control signaling.
	
\vspace{-1em}
\subsection{Problem Formulation} \label{subsec:prob formulation}
	Denote by $P_r, P_b$ and $P_{u,k}$ the maximum transmission powers of the UAV, the BS and each UE $k$, respectively. 
	By \eqref{snr:ul:BS}, \eqref{snr:dl:UE}, \eqref{sRate} and \eqref{snr:ctrl:dl}, 
	we consider the following joint UAV positioning and power control (JPPC) problem 
	\begin{subequations} \label{mulUE_opt:1}
		\begin{align} 
		\max_{\substack{ \xb_r, ~ \pb_b,\pb_r^{\rm U},\pb_r^{\rm D},\pb_u \geq  0,\\  p_{c} \ge 0} }~ & R_s(\xb_r,  \pb_b,\pb_r^{\rm U},\pb_r^{\rm D},\pb_u) \\ \label{cons:uav}
		\text{s.t.}~~ 
		&    \oneb^T \pb_{r}^{U} +\oneb^T  \pb_{r}^{\rm D}  + p_{c} \le P_r,\\ \label{cons:bs}
		& \oneb^T \pb_{b} + p_{c} \le P_b,\\ \label{cons:ue}		
		& p_{u,k} \le  P_{u,k}, k \in \mathcal{K}, \\ 
		& {\rm SNR}_c \ge  \gamma_c,\label{cons:ctrl}
		\end{align} 
	\end{subequations}
	where $\mathbf{1}$ is the all-one vector, and $\gamma_c$ is the SNR requirement of the downlink and uplink control signaling. The constraints (\ref{cons:uav}) and (\ref{cons:bs}) are the total transmission power constraints at the UAV and the BS, respectively; \eqref{cons:ue} constrains the maximum transmission power of each UE $k$.
	\begin{property} \label{lemma: p_{u,k}} 
		All constraints \eqref{cons:uav} to \eqref{cons:ctrl} of problem \eqref{mulUE_opt:1} hold with equality at the optimum.
	\end{property}
	
	{\emph{Proof}:} 
	It is easy to verify that ${\rm SNR}_k^{\rm U}$ in \eqref{snr:ul:BS} is an increasing function of $p_{u,k}$ and $p_{r,k}^{\rm U}$, respectively; similary, ${\rm SNR}_k^{\rm D}$ in \eqref{snr:dl:UE} is an increasing function of $p_{b,k}$ and $p_{r,k}^{\rm D}$, respectively. Thus, constraints \eqref{cons:uav} to \eqref{cons:ue} must hold with equality at the optimum.
	If \eqref{cons:ctrl} holds with strict inequality at the optimum, then one can reduce $p_c$ and it makes \eqref{cons:uav} to \eqref{cons:ue} inactive. 
	Then, either $p_{b,k}$,  $p_{r,k}^{\rm U}$ or $p_{r,k}^{\rm D}$ can be further increased to improve the sum rate. 
	As a result,  \eqref{cons:ctrl} must also hold with equality at the optimum. 
	\hfill $\blacksquare$
	
	By Property \ref{lemma: p_{u,k}}, we obtain the optimal $p_{u,k} = P_{u,k}, \forall k\in \Kc,$ and  $p_{c}= \gamma_c \xi^{-1} \|\xb_r-\bb \|^2$ for problem \eqref{mulUE_opt:1}.
	Thus, problem (\ref{mulUE_opt:1}) can be simplified as
	\begin{subequations} \label{mulUE_opt:noPu}
		\begin{align} 
		\max_{\substack{  \xb_r, ~\pb \geq 0}}~ & R_s(\xb_r, \pb) 
		\\ \label{cons:2:uav}
		\text{s.t.}~ &  \oneb^T \pb_{r}^{U} +\oneb^T  \pb_{r}^{\rm D}   + \frac{\gamma_c}{\xi} \|\xb_r-\bb \|^2 \le P_r,
		\\ \label{cons:2:bs}
		& \oneb^T \pb_{b} + \frac{\gamma_c}{\xi} \|\xb_r-\bb \|^2 \le P_b,
		\end{align} 
	\end{subequations}
	where $\pb \triangleq (({\pb_{r}^{\rm U}})^T,({\pb_{r}^{\rm D}})^T,\pb_{b}^T)^T$, and, with a slight of abuse of notation,
	\begin{align}
	R_s(\xb_r, \pb)  & \triangleq   \sum_{k=1}^K \bigg( R_k^{\rm U}(\xb_r, p_{r,k}^{\rm U})  + R_k^{\rm D}(\xb_r, p_{b.k},p_{r,k}^{\rm D}) \bigg), \label{sum rate}\\
	R_k^{\rm U}(\xb_r, p_{r,k}^{\rm U}) &=\frac{W}{2} \text{log}\bigg(1+  
	\frac{  p_{r,k}^{\rm U}  P_{u,k} d_{kr}^{-2} d_{rb}^{-2} \xi  }{p_{r,k}^{\rm U} d_{rb}^{-2} +  P_{u,k} d_{kr}^{-2} + \xi^{-1}} \bigg), \label{U rate} \\
	R_k^{\rm D}(\xb_r, p_{b.k},p_{r,k}^{\rm D}) & =
	\frac{W}{2} 
	\text{log}\bigg(1+ \frac{  p_{r,k}^{\rm D}  p_{b,k} d_{rb}^{-2} d_{kr}^{-2} \xi }{ p_{r,k}^{\rm D} d_{kr}^{-2} +   p_{b,k} d_{rb}^{-2} + \xi^{-1}}\bigg).\label{D rate}
	\end{align}
\section{UAV Positioning and Power Control: Single UE Case }\label{sec:sigUE}
	To gain more insights, let us first study a special instance of problem \eqref{mulUE_opt:noPu} with only one UE ($K=1$).
	For the signle UE case, problem \eqref{mulUE_opt:noPu} reduces to
	\begin{subequations} \label{sigUE_opt:uldl}
		\begin{align}
		\max_{\substack{  \xb_r, ~\pb \geq 0}}~ & R_s(\xb_r, \pb)  \\
		\text{s.t.}~ &  (p_{r}^{U} + p_{r}^{\rm D} ) + \frac{\gamma_c}{\xi} \|\xb_r-\bb \|^2 \le P_r,
		\\ 
		& p_{b} + \frac{\gamma_c}{\xi} \|\xb_r-\bb \|^2 \le P_b,
		\end{align} 
	\end{subequations}
	where 
	\begin{align}  \label{sigUE_opt:uldl:obj1}
	R_s(\xb_r, \pb)  =  \frac{W}{2} \bigg( &\text{log}(1+  
	\frac{  p_{r}^{\rm U}  P_{u} \xi  }{p_{r}^{\rm U} d_{ur}^2 +  P_{u} d_{rb}^2+ \xi^{-1}d_{ur}^{2} d_{rb}^{2}   } ) 
	\notag \\
	&+ \text{log}(1+ \frac{  p_{r}^{\rm D}  p_{b}  \xi }{ p_{r}^{\rm D} d_{rb}^{2} +   p_{b} d_{ur}^{2} + \xi^{-1}d_{rb}^{2} d_{ur}^{2}  }) \bigg).
	\end{align}
	Here, the subscript $k$ of all variables is removed for notation simplicity; besides, each $d_{kr}$ is replaced by $d_{ur}=\|\xb_r -\ub\|$ in which $\ub$ is the 3D location of the UE.

	It is not surprising to see that the following statement is true.
	\begin{property} \label{lemma: x_r} 
		When projected onto the x-y plane, the optimal UAV position is on the line segment connecting the BS and the UE.
	\end{property}
	
	{\emph{Proof}:}  The proof is presented in Appendix \ref{proof of property 2}.

	By Property \ref{lemma: x_r}, we can write $\xb_r=\ub + \alpha \ssb + h \eb_z$, where  $\ssb=\frac{\bb - \ub}{\|\bb - \ub\|}$,  $\eb_z=(0,0,1)^T$ and $0\geq  \alpha \leq M\triangleq \|\bb-\ub\|$. By this expression and Property \ref{lemma: p_{u,k}}, we have 
	\begin{align}\label{dur}
	d_{ur}^2&=\|\xb_r- \ub\|^2=  \alpha^2+ h^2, \\
	d_{rb}^2&=\|\xb_r- \bb\|^2=  (M-\alpha)^2+ h^2,\label{drb} \\
	p_b &= P_b - \frac{\gamma_c}{\xi} ((M-\alpha)^2 + h^2). \label{pb}
	\end{align}
	Thus, problem \eqref{sigUE_opt:uldl} is equivalent to the following problem
	\begin{align}
	&\max_{\substack{ 0 \le \alpha \le M}}~ 
	\left\lbrace
	\begin{array}{ll}
	&\displaystyle \max_{\substack{ p_r^{\rm U} \ge 0, p_r^{\rm D}\ge0}}  R_s\left(\alpha, p_r^{\rm U}, p_r^{\rm D}\right) \\
	&~~~~~~~\text{s.t.}~	 p_{r}^{\rm U} + p_{r}^{\rm D} \le P_r -\frac{\gamma_c}{\xi} ((M-\alpha)^2 + h^2)
	\end{array} \right\rbrace
	\label{sigUE_opt:uldl:obj}	
	\\
	= & \max_{\substack{ 0 \le \alpha \le M}} R^\star_s\left(\alpha\right)  \label{sigUE_opt:uldl:obj2}	
	\end{align}
	where $R_s\left(\alpha, p_r^{\rm U}, p_r^{\rm D}\right)$ is obtained by substituting \eqref{dur}, \eqref{drb} and \eqref{pb} into \eqref{sigUE_opt:uldl:obj1}, and $R^\star_s\left(\alpha\right)$ denotes the optimal sum rate of the inner problem in \eqref{sigUE_opt:uldl:obj} with a given value of $\alpha$. 
	It is worth noting that, while problem \eqref{sigUE_opt:uldl:obj} is not a convex problem, the inner problem with a given value of $\alpha$ is a convex problem (since $ R_s\left(\alpha, p_r^{\rm U}, p_r^{\rm D}\right)$ is a concave function for $p_r^{\rm U}, p_r^{\rm D}\geq 0$), which 
	can be efficiently solved.
	Therefore, one can globally solve  problem \eqref{sigUE_opt:uldl} by searching the optimal value of $\alpha \in [0,M]$ in \eqref{sigUE_opt:uldl:obj2}.
\section{UAV Positioning and Power Control: Multiple User Case }\label{sec:mulUE}
	In this section, we study efficient algorithms to solve the UAV positioning and power control problem \eqref{mulUE_opt:noPu} with multiple UEs. 
	Unlike the single user case, \eqref{mulUE_opt:noPu} is much more challenging to solve due to the non-concave objective function.
	Our aim is to develop computationally efficient algorithms for \eqref{mulUE_opt:noPu}. 
	Specifically, the proposed approach is based on the successive convex approximation (SCA) method \cite{SCA_1,Razaviyayn_parallelBCD14,SCA}, where
	one obtains a suboptimal solution by solving a sequence of convex surrogate problems. 
	For our problem \eqref{mulUE_opt:noPu}, since the constraints \eqref{cons:2:uav} and \eqref{cons:2:bs} are both convex, we need to find a proper concave surrogate function for the non-concave sum rate function $R_s(\xb_r, \pb)$.
	Next, we propose such a concave surrogate function that is amenable for fast SCA convergence.

\vspace{-1em}
\subsection{Proposed SCA Algorithm}  \label{sec:mulUE:surro1}
	Let us present a concave surrogate function for $R_s(\xb_r, \pb)$ in \eqref{sum rate}.
	Let $(\bar \pb_{b}, \bar \pb_{r}^{\rm D}, \bar{\pb}_{r}^{\rm U}, \bar\xb_r)$
	be a feasible point to problem \eqref{mulUE_opt:noPu}.	Define
	\begin{subequations}
		\begin{align}
		\bar{\db}_{rb} &\triangleq \bar\xb_r - \bb,~{\db}_{rb} \triangleq \xb_r - \bb\\
		\bar{\db}_{kr} &\triangleq \bar\xb_r - \ub_k,~{\db}_{kr} \triangleq \xb_r - \ub_k,~ k \in \mathcal{K},
		\end{align}
	\end{subequations}
	and
	\begin{align}
	\bar{I}_k^{\rm D}   &\triangleq  \xi^{-1} \frac{\|\bar \db_{rb}\|^2}{\bar p_{b,k}} + \xi^{-1} \frac{\|\bar \db_{kr}\|^2}{\bar p_{r,k}^{\rm D}} + \xi^{-2} \frac{\|\bar \db_{rb}\|^2}{\bar p_{b,k}}\frac{\|\bar \db_{kr}\|^2}{\bar p_{r,k}^{\rm D}},   \label{IkD}\\
	\bar{I}_k^{\rm U} &\triangleq  \xi^{-1} \frac{\|\bar\db_{kr}\|^2}{\bar{p}_{u,k}} + \xi^{-1} \frac{\|\bar\db_{rb}\|^2}{\bar{p}_{r,k}^{\rm U}} + \xi^{-2} \frac{\|\bar\db_{kr}\|^2}{\bar{p}_{u,k}}\frac{\|\bar\db_{rb}\|^2}{\bar{p}_{r,k}^{\rm U}}, \label{IkU}
	\end{align}
	for all $ k \in \mathcal{K}$.
	\begin{prop}  \label{prop surrogate func} 
		The function 
		\begin{align}\label{surrogate 1}
		\bar {R}_s\left( \xb_r, {\pb} \right)  \triangleq \frac{W}{2} \sum_{k=1}^K (\bar R_k^{\rm D} ( \xb_r, p_{b,k},  p_{r,k}^{\rm D}) + \bar R_k^{\rm U} ( \xb_r, p_{r,k}^{\rm U})),
		\end{align} 
		where $\bar R_k^{\rm D} ( \xb_r, p_{b,k}, p_{r,k}^{\rm D})$ and
		$\bar R_k^{\rm U} (\xb_r, p_{r,k}^{\rm U})$ are respectively given in \eqref{RkDs} and \eqref{RkUs} at the bottom of next page,
		is a concave function and is a locally tight lower bound of the sum rate function \eqref{sum rate}, i.e.,
		\begin{subequations}\label{locally tight D}
			\begin{align}
			&R_s( \bar\xb_r, \bar \pb) = \bar R_s ( \bar\xb_r, \bar \pb),\label{surrogate func D tight}\\
			&R_s ( \xb_r, \pb) \geq \bar R_s ( \xb_r,  \pb),~  \label{surrogate func D}
			\end{align}
		\end{subequations}
		for all feasible $(\xb_r, \pb)$
		\vspace{-0cm}
	\end{prop}
	
	{\emph{Proof}:} 
	The derivations of \eqref{RkDs} and \eqref{RkUs} are technical. They are obtained by carefully examining the function structure and applying the first-order Taylor lower bound of convex components in the rate functions.
	The details are given in Appendix \ref{proof of surrogate functions}.
	\hfill $\blacksquare$
	
	\begin{figure*}[!b]
		\normalsize
		\hrulefill
		\begin{align}	
		&\bar R_k^{\rm D} ( \xb_r,  p_{b,k},  p_{r,k}^{\rm D}) \triangleq \text{log}\left( 1 + \xi^{-1} \left(   \frac{2\bar\db_{rb}^T}{\bar p_{b,k}}\db_{rb} - \frac{\|\bar\db_{rb}\|^2}{\bar p_{b,k}^2}  p_{b,k}\right) \right) + \text{log}\left( 1 + \xi^{-1} \left(  \frac{2\bar\db_{kr}^T}{\bar p_{r,k}^{\rm D}}\db_{kr}  - \frac{\|\bar\db_{kr}\|^2}{(\bar p_{r,k}^{\rm D})^2} p_{r,k}^{\rm D}\right) \right)  \notag  \\  
		&~~~~~~~~~~~~~~~~~~~~~~~  - \text{log}(\bar I_k^{\rm D}) + 1 - \frac{1}{\bar I_k^{\rm D} \xi} \left( \frac{\|\db_{rb}\|^2}{p_{b,k}} +  \frac{\|\db_{kr}\|^2}{p_{r,k}^{\rm D}} \right) - \frac{1}{2 \bar I_k^{\rm D} \xi^2}  \left(  \frac{\|\db_{rb}\|^2}{p_{b,k}} + \frac{\|\db_{kr}\|^2}{p_{r,k}^{\rm D}} \right) ^2  \notag  \\  
		&~~~~~~~~~~~~~~~~~~~~~~~+ \frac{1}{2 \bar I_k^{\rm D} \xi^2} \bigg[  \left( \frac{4\|\bar\db_{rb}\|^2 \bar\db_{rb}^T \db_{rb}}{\bar p_{b,k}^2}- \frac{2 (\|\bar\db_{rb}\|^2)^2 p_{b,k}}{\bar p_{b,k}^3}  - \frac{(\|\bar\db_{rb}\|^2)^2}{\bar p_{b,k}^2} \right) \notag \\
		&~~~~~~~~~~~~~~~~~~~~~~~
		~~~~~~~~~~~~~+\left(   \frac{4\|\bar\db_{kr}\|^2 \bar\db_{kr}^T \db_{kr}}{(\bar p_{r,k}^{\rm D})^2}- \frac{2 (\|\bar\db_{kr}\|^2)^2 p_{r,k}^{\rm D}}{(\bar p_{r,k}^{\rm D})^3}  - \frac{(\|\bar\db_{kr}\|^2)^2}{(\bar p_{r,k}^{\rm D})^2}\right)   \bigg], \label{RkDs} \\
		& \bar {R}_k^{\rm U} ( \xb_r, p_{r,k}^{\rm U})
		\triangleq \text{log}\left( 1 + \xi^{-1} \left(  \frac{2\bar\db_{kr}^T}{P_{u,k}}\db_{kr} - \frac{\|\bar\db_{kr}\|^2}{ P_{u,k}}  \right)  \right) + \text{log}\left( 1 + \xi^{-1} \left(  \frac{2\bar\db_{rb}^T}{\bar p_{r,k}^{\rm U}}\db_{rb} - \frac{\|\bar\db_{rb}\|^2}{(\bar p_{r,k}^{\rm U})^2}  p_{r,k}^{\rm U} \right)  \right)  \notag \\
		&~~~~~~~~~~~~~~~~~~~~~~~- \text{log}(\bar I_k^{\rm U}) + 1- \frac{1}{\bar I_k^{\rm U} \xi} \left( \frac{\|\db_{kr}\|^2}{P_{u,k}} +  \frac{\|\db_{rb}\|^2}{p_{r,k}^{\rm U}} \right) - \frac{1}{2 \bar I_k^{\rm U} \xi^2}  \left( \frac{\|\db_{kr}\|^2}{P_{u,k}} + \frac{\|\db_{rb}\|^2}{p_{r,k}^{\rm U}} \right) ^2 \notag \\
		& ~~~~~~~~~~~~~~~~~~~~~~~
		+ \frac{1}{2 \bar I_k^{\rm U} \xi^2} \bigg[   \left( \frac{4\|\bar\db_{kr}\|^2 \bar\db_{kr}^T \db_{kr}}{P_{u,k}^2}  - \frac{3(\|\bar\db_{kr}\|^2)^2}{ P_{u,k}^2} \right) \notag \\
		&~~~~~~~~~~~~~~~~~~~~~~~
		~~~~~~~~~~~~~	+\left( \frac{4\|\bar\db_{rb}\|^2 \bar\db_{rb}^T \db_{rb}}{(\bar p_{r,k}^{\rm U})^2}- \frac{2 (\|\bar\db_{rb}\|^2)^2 p_{r,k}^{\rm U}}{(\bar p_{r,k}^{\rm U})^3}  - \frac{(\|\bar\db_{rb}\|^2)^2}{(\bar p_{r,k}^{\rm U})^2} \right)  \bigg]. \label{RkUs}
		\end{align}	
		\vspace{-0.9pt}
	\end{figure*}

	By replacing the objective function of \eqref{mulUE_opt:noPu}
	by \eqref{surrogate 1},
	we obtain the following convex optimization problem
	\begin{subequations} \label{mulUE_opt:2:app}
		\begin{align} \label{mulUE_opt:2:app:obj}
		\max_{\substack{\xb_r,~{\pb} \ge 0 }} &~~ \bar {R}_s\left( \xb_r, {\pb} \right) \\ 
		\text{s.t.}~ &  \oneb^T \pb_{r}^{U} +\oneb^T  \pb_{r}^{\rm D}    + \frac{\gamma_c}{\xi}\|\xb_r - \bb\|^2 \le P_r, 
		\label{mulUE_opt:2:app:obj C1} \\
		& \oneb^T \pb_{b}+ \frac{\gamma_c}{\xi}\|\xb_r - \bb\|^2 \le P_b. \label{mulUE_opt:2:app:obj C2}
		\end{align}
	\end{subequations}
	The proposed SCA algorithm for solving problem \eqref{mulUE_opt:noPu} then iteraively solves \eqref{mulUE_opt:2:app}
	with a given feasible point $(\bar \pb, \bar\xb_r)$ obtained in the previous iteration, as shown in Algorithm \ref{alg:mulULDL}. Since the constraint set of problem \eqref{mulUE_opt:noPu} is compact and convex, according to \cite[Corollary 1]{BSUMM}, it can be shown that the variables $(\pb,\xb_r)$ yielded by Algorithm \ref{alg:mulULDL} converges to a stationary point of problem \eqref{mulUE_opt:noPu} as the iteration number goes to infinity. 	
	\begin{algorithm}[t]
		\caption{Proposed SCA Algorithm for Problem (\ref{mulUE_opt:noPu})}
		\begin{algorithmic}[1] \label{alg:mulULDL}
			\STATE {\bf Given}  an initial point $(\pb^0, \xb_r^0)$ that is feasible to problem (\ref{mulUE_opt:noPu}); Set $i=0$.
			\REPEAT
			\STATE {Update} $(\bar \pb, \bar\xb_r)$ by $(\pb^i, \xb_r^i)$.
			\STATE {Solve} problem (\ref{mulUE_opt:2:app}) and obtain the optimal solution $(\pb^{i+1}, \xb_r^{i+1})$.
			\STATE  $i\leftarrow i+1$.
			\UNTIL  $R_s(\xb_r^{i}, \pb^{i})-R_s(\xb_r^{i-1}, \pb^{i-1}) \leq \epsilon_0$
		\end{algorithmic}
		\label{alg: SCA}
	\end{algorithm}		
	\begin{remark}\rm \label{rmk1 inner subp}
		It is worthwhile to mention that, except for using the off-the-shelf convex solvers such as CVX \cite{CVX} to solve \eqref{mulUE_opt:2:app}, it would be more efficient to develop a customized algorithm. For example, because the Slater's condition holds for  \eqref{mulUE_opt:2:app}, one may consider the Lagrange dual problem of \eqref{mulUE_opt:2:app}, i.e.,
		\begin{equation} \label{dual_opt}
		\max_{\substack{ \lambda \ge 0, \mu \ge 0 }} \left\lbrace \min_{\substack{\xb_r, {\pb}\geq 0} }  \mathcal{L}_1 \left( {\mathbf{x}}_r, {\pb}, \lambda, \mu\right)\right\rbrace,
		\end{equation}  
		where 	
		\begin{align}  \label{Lag:mulUE_pot:2}
		\mathcal{L}_1\left( {\mathbf{x}}_r, {\pb}, \lambda, \mu  \right) = - \bar {R}_s\left( \xb_r, {\pb} \right)    &+ \lambda \left(  \oneb^T \pb_{r}^{U} +\oneb^T  \pb_{r}^{\rm D}    + \frac{\gamma_c}{\xi}\|\xb_r - \bb\|^2 - P_r \right)   \notag \\ 
		&+ \mu \left( \oneb^T \pb_{b}+ \frac{\gamma_c}{\xi}\|\xb_r - \bb\|^2 - P_b \right)  
		\end{align}
		is the Lagrangian function, and $\lambda\geq 0$ and $\mu\geq 0$ are the dual variables associated with \eqref{mulUE_opt:2:app:obj C1}  and \eqref{mulUE_opt:2:app:obj C2}, respectively. The dual subgradient ascent (DSA) method \cite{Boydsubgradient} can be applied to \eqref{dual_opt} while the inner minimization problem 
		$\min_{{\xb_r, {\pb}\geq 0} }  \mathcal{L} ( {\mathbf{x}}_r, {\pb}, \lambda, \mu)$ can be solved by applying the gradient projection (GP) method \cite{BK:Bersekas_NLP}.
		Since $\mathcal{L}_1\left( {\mathbf{x}}_r, {\pb}, \lambda, \mu  \right)$ has a separable strucutre (it is a summation and each of the terms involves variables of either one UE or the UAV only),
		the GP method for the inner minimization problem can inherently be implemented in a fully parallel manner.
		The resultant algorithm is therefore more time efficient than the general-purpose solvers. 
	\end{remark}
	
\vspace{-1em}
\subsection{Comparison with the Surrogate Function in \cite{ShenChang018}}  \label{sec:mulUE:surro2} 
	It is important to notice that the locally tight surrogate functions presented in Proposition \ref{prop surrogate func} are simply one of the choices for SCA optimization, and a different surrgorate function may be obtained by another approach.
	From a theoretical point of view, as long as the surrogate function is a locally tight lower bound, i.e., satisfies \eqref{locally tight D}, convergence of the SCA algorithm is guaranteed. Nevertheless, different surrograte functions may result in quite different convergence behavior. In accordance with \cite[Theorem 3]{Razaviyayn_parallelBCD14}, the iteration complexity of the SCA algorithm is in the order of $\mathcal{O}(\frac{L^2}{\epsilon})$, where $\epsilon$ is a solution accuracy and $L$ is the gradient Lipschitz constant of the employed surrogate function. 
	The constant $L$ represents the curvature and is also the spectral radius of the Hessian matrix of the surrogate function provided that it is twice differentiable.
	
	In this subsection, we aim to demonstrate that the proposed surrogate functions in Proposition \ref{prop surrogate func} is good in the sense that it has a faster convergence behavior than a surrogate function that is deduced following the idea in a recent work \cite{ShenChang018}.
	%
	In particular, as we show in Appendix \ref{app:cmp_surrogate12}, by following a similar method as in \cite[Eqn. (20)]{ShenChang018},
	one can obtain 
	\begin{align}\label{surrogate 2}
	\hat {R}_s\left( \xb_r, {\ab} \right)  \triangleq \frac{W}{2} \sum_{k=1}^K (\hat R_k^{\rm D} ( \xb_r, a_{b,k},  a_{r,k}^{\rm D}) + \hat R_k^{\rm U} ( \xb_r, a_{r,k}^{\rm U})),
	\end{align}
	as another concave and locally tight lower bound for $R_s( \xb_r, \pb)$ in \eqref{sum rate}.
	Here
	\begin{align}\label{surrogate D2}
	\hat R_k^{\rm D}( \xb_r, a_{b,k}, a_{r,k}^{\rm D}) &=
	\text{log} \left( 1 + \xi \frac{2 \bar a_{r,k}^{\rm D} }{ \|\bar \db_{kr}\|^2 } a_{r,k}^{\rm D} - \xi \frac{ (\bar{a}_{r,k}^{\rm D})^2 }{ (\|\bar \db_{kr}\|^2)^2} \|\db_{kr} \|^2 \right) \notag \\
	&~~~+ \text{log} \left( 1 + \xi \frac{2 \bar a_{b,k} }{\|\bar \db_{rb}\|^2 } a_{b,k} - \xi \frac{ (\bar a_{b,k})^2 }{(\|\bar \db_{rb}\|^2)^2 } \|\db_{rb} \|^2 \right) 
	- \text{log} \left( 1 + \bar{J}_k^{\rm D}\right) + \frac{\bar{J}_k^{\rm D}}{1 + \bar{J}_k^{\rm D}}  \notag \\
	& ~~~ - \frac{\xi}{1 + \bar{J}_k^{\rm D}} \left( \frac{({a_{r,k}^{\rm D}})^2}{\| {\ub}_k\|^2- \|\bar \xb_r\|^2 + 2 (\bar \db_{kr})^T \xb_r }  + \frac{a_{b,k}^2}{\| {\bb}\|^2 - \|\bar \xb_r\|^2 + 2 (\bar \db_{rb})^T \xb_r} \right),
	\\
	\hat R_k^{\rm U} \left( \xb_r, a_{r,k}^{\rm U} \right) &=
	\text{log} \left( 1 + \xi \frac{2 \bar a_{r,k}^{\rm U} }{\|\bar \db_{rb}\|^2 } a_{r,k}^{\rm U} - \xi \frac{ (\bar{a}_{r,k}^{\rm U})^2 }{(\|\bar \db_{rb}\|^2)^2 } \|\db_{rb} \|^2 \right)  \notag \\
	&~~~+ \text{log} \left( 1 + \xi \frac{2 P_{u,k} }{\|\bar \db_{kr}\|^2 } - \xi \frac{ P_{u,k} }{(\|\bar \db_{kr}\|^2)^2 } \|\db_{kr} \|^2 \right) 
	- \text{log} \left( 1 + \bar{J}_k^{\rm U}\right) + \frac{\bar{J}_k^{\rm U}}{1 + \bar{J}_k^{\rm U}}  \notag \\
	& ~~~ - \frac{\xi}{1 + \bar{J}_k^{\rm U}} \left( \frac{({a_{r,k}^{\rm U}})^2}{\| {\bb}\|^2 - \|\bar \xb_r\|^2 + 2 (\bar \db_{rb})^T \xb_r}  + \frac{P_{u,k}}{ \| {\ub}_k\|^2- \|\bar \xb_r\|^2 + 2 (\bar \db_{kr})^T \xb_r} \right).  \label{surrogate U2}
	\end{align}	
	where $a_{b,k}=\sqrt{p_{b,k}}, a_{r,k}^{\rm D}=\sqrt{p_{r,k}^{\rm D}}$ and $a_{r,k}^{\rm U}=\sqrt{p_{r,k}^{\rm U}}$; $\bar a_{b,k}=\sqrt{\bar p_{b,k}}, \bar a_{r,k}^{\rm D}=\sqrt{\bar p_{r,k}^{\rm D}}$ and $\bar a_{r,k}^{\rm U}=\sqrt{\bar p_{r,k}^{\rm U}}$;  
	and
	\begin{align}\label{bar J}
	\bar J_k^{\rm D} = \xi \bigg(\frac{(\bar a_{r,k}^{\rm D})^2}{\|\ub_k -\bar \xb_r\|^2} + \frac{\bar a_{b,k}^2}{\|\bar \xb_r - \bb\|^2}\bigg),~\bar{J}_k^{\rm U}=\xi \bigg(\frac{(\bar a_{r,k}^{\rm U})^2}{\|\bar \xb_r - \bb\|^2} + \frac{P_{u,k}}{\|\ub_k -\bar \xb_r\|^2}\bigg). 
	\end{align}
	
	Next we compare the two surrogate functions in \eqref{surrogate 1} and \eqref{surrogate 2} analytically and numerically.
	For ease of illustration, we focus on the UAV position variable $\xb_r$ only and assume that the power variables $\pb$ are fixed at the given value $\bar \pb$. As mentioned, by \cite[Theorem 3]{Razaviyayn_parallelBCD14}, the iteration complexity of the SCA algorithm to reach a stationary point is in the order of $\mathcal{O}(\frac{L^2}{\epsilon})$.
	If a surrogate function has a larger value of $L$, the surrogate function has a larger curvature and thus the SCA algorithm would progress slowly.
	The following result compares the curvature of the two surrogate functions at the given point $(\bar \xb_r, \bar \pb)$.
	{	\begin{prop}
			\label{prop: curvature}
			Consider the surrogate functions $\bar R_s ( \xb_r, \bar \pb)$ in \eqref{surrogate 1} and $\hat R_s ( \xb_r, \bar \pb)$ in \eqref{surrogate 2}. When $\xi$ is large, the spectral radius of the Hessian matrices $\nabla^2_{\xb_r} \bar R_s ( \xb_r, \bar \pb))$ and $\nabla^2_{\xb_r} \hat R_s ( \xb_r, \bar \pb))$ satisfy
			\begin{align}
			&|\lambda_{\max}(\nabla^2_{\xb_r} \bar R_s ( \xb_r, \bar \pb))| \leq \mathcal{O}(\xi^{-1}) + c, \\
			&|\lambda_{\max}(\nabla^2_{\xb_r} \hat  R_s ( \xb_r, \bar \pb))| \leq \mathcal{O}(1),
			\end{align}
			respectively, where $c>0$ is a constant.
			Moreover, when $\xb_r=\bar \xb_r$ and $\xi \rightarrow \infty$,			
			\begin{align}\label{comp of curvature}
			|\lambda_{\max}(\nabla^2_{\xb_r} \hat R_s ( \bar \xb_r, \bar \pb))| > |\lambda_{\max}(\nabla^2_{\xb_r} \bar R_s ( \xb_r, \bar \pb))| .
			\end{align}
		\end{prop} 
		
		\emph{Proof:}  The result is obtained by deriving upper and lower bounds of the Hessian matrices of the surrogate functions; details are given in Appendix \ref{proof:Hessian_cmp}.
		\hfill $\blacksquare$ 
	}
	
\begin{figure}[!t]	\centering
		\subfigure[3D curve] {\label{fig:convergence_a} \includegraphics[width=0.47\linewidth]{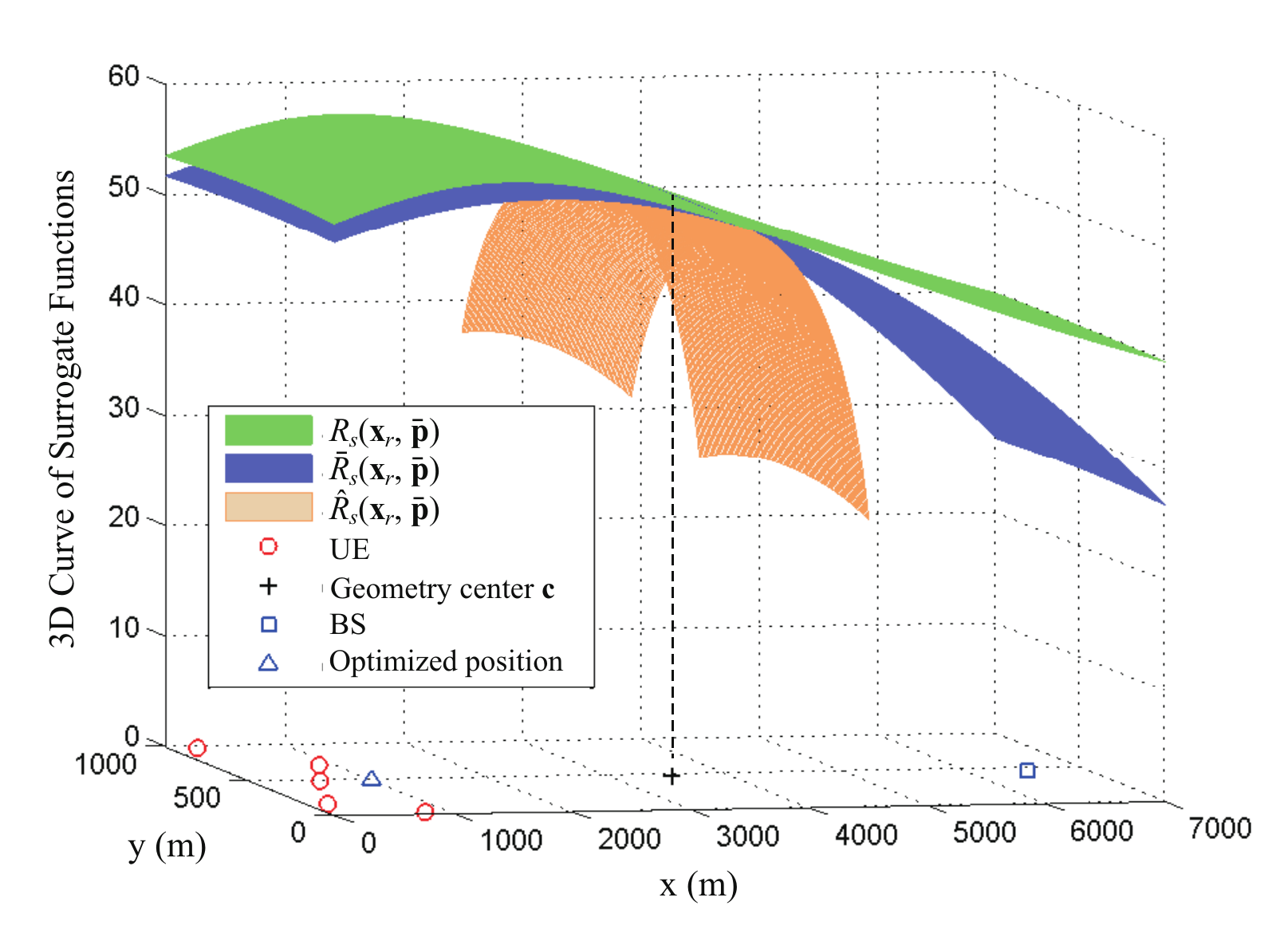} }
		\subfigure[Side view along the x-axis] { \label{fig:convergence_b}  \includegraphics[width=0.47\linewidth]{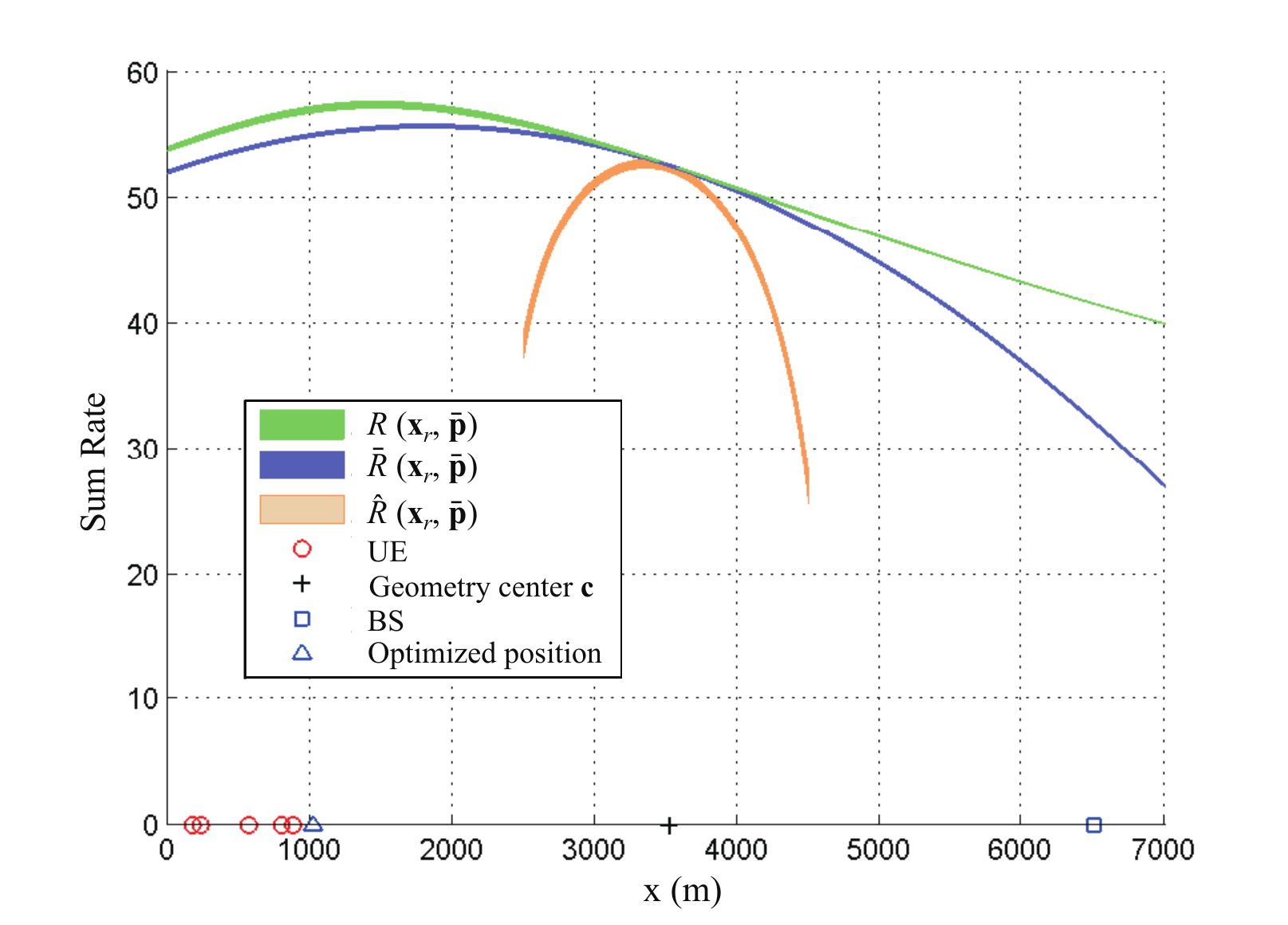} }
		\caption{Illustration of the sum rate function $R_s ( \xb_r, \bar \pb)$ and the surrogate functions $\bar R_s ( \xb_r, \bar \pb)$ and $\hat R_s (\xb_r, \bar \pb)$
			with respect to the UAV position $\xb_r$ for a scenario with $K=5$ UEs; 
			$\bar R_s ( \xb_r, \bar \pb)$ and $\hat R_s ( \xb_r, \bar \pb)$ are obtained by setting $\bar \xb_r$ equal to the geometry center point and the power values are set to $\bar p_{r,k}^{\rm D}=\bar p_{r,k}^{\rm U}=P_r/(2K)$ and $p_{b,k}=P_b/K$ for all $k\in \Kc$.}
		\label{fig:convergence}
		\vspace{-0.5cm}
	\end{figure}

	As $\xi$ is typically a large number\footnote{ For $\beta = -40 ~\text{dB}$ and  $\sigma^2=-169~\text{dBm/Hz} \times 10~\text{MHz}=-99$ dBm, $\xi $ is approximately $10^9$.}, Proposition \ref{prop: curvature} shows that the curvature of the proposed surrogate function in \eqref{surrogate 1}
	can be smaller than that of the surrogate function in \eqref{surrogate 2} at the approximation point.    
	While Proposition  \ref{prop: curvature} gives only a limited claim, the curvature difference between the two surrogate functions can actually be large numerically.
	To demonstrate this, we draw in Fig. \ref{fig:convergence} (3D curve in Fig. \ref{fig:convergence}(a) and side view along the x-axis in Fig. \ref{fig:convergence}(b)) the sum rate function $R_s ( \xb_r, \bar \pb)$ and the surrogate functions $\bar R_s ( \xb_r, \bar \pb)$ and $\hat R_s (\xb_r, \bar \pb)$
	with respect to the UAV position $\xb_r$ for a scenario with $K=5$ UEs.
	The  simulation setting is the same as that in Section \ref{sec:simu}.
	First of all, one can see that the sum rate function $R_s ( \xb_r, \bar \pb)$ is non-concave whereas the two surrogate functions $\bar R_s ( \xb_r, \bar \pb)$ and $\hat R_s (\xb_r, \bar \pb)$ are concave and lower bounds of $R_s ( \xb_r, \bar \pb)$.
	Secondly, at the given point of $\bar \xb_r$ which is equal to the geometry center $\mathbf{c} \triangleq (\frac{1}{K} \sum_{k=1}^K \ub_k + \bb)/2$, the proposed surrogate function  
	$\bar R_s ( \xb_r, \bar \pb)$ is much less curvy than the surrogate function $\hat R_s (\xb_r, \bar \pb)$.
	Besides, comparing to $\hat R_s (\xb_r, \bar \pb)$, the maximum function value of $\bar R_s ( \xb_r, \bar \pb)$ is closer to that of $R_s ( \xb_r, \bar \pb)$.
	Therefore, in accordance with \cite[Theorem 3]{Razaviyayn_parallelBCD14}, one can anticipate that the SCA algorithm using $\bar R_s ( \xb_r, \pb)$ would exhibit a faster convergence behavior.
	
\begin{figure}[!t]
		\centering
		\includegraphics[width=0.5\linewidth]{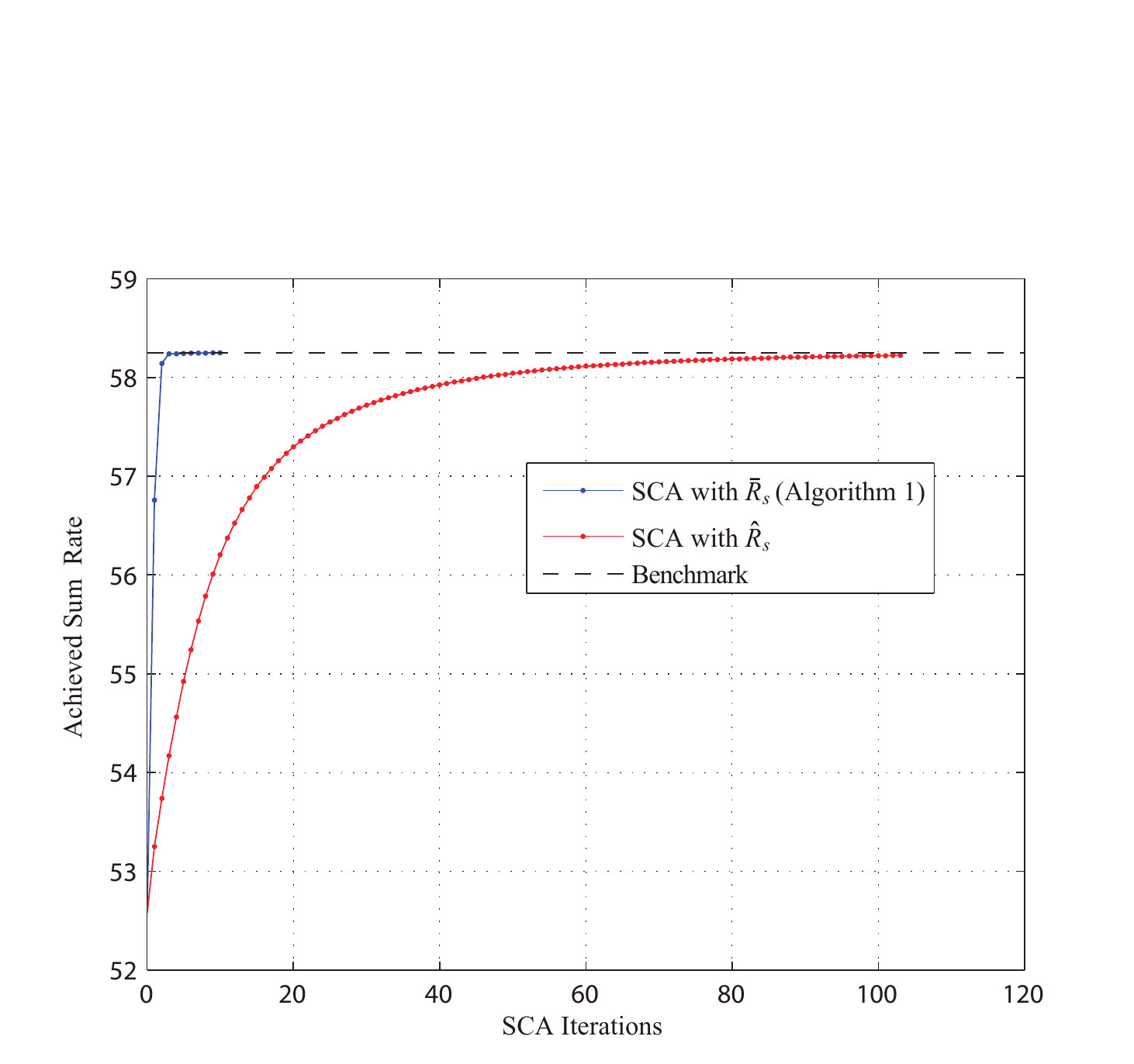}
		\vspace{-0.3cm}
		\caption{Convergence curves of the proposed SCA algorithm using surrogate functions $\bar R_s ( \xb_r, \pb)$ in \eqref{surrogate 1} and $\hat R_s ( \xb_r, \pb)$ in \eqref{surrogate 2}, respectively.}
		\label{fig:surrogate_cvg_cmp}
		\vspace{-0.8cm}
	\end{figure}

	In Fig. \ref{fig:surrogate_cvg_cmp}, we further show the convergence curves (achieved sum rate versus iteration number) of the SCA algorithms using surrogate functions $\bar R_s ( \xb_r, \bar \pb)$ and $\hat R_s (\xb_r, \bar \pb)$, respectively.
	Once can observe from the figure that the two converge curves are drastically different -- the SCA algorithm using $\bar R_s ( \xb_r, \bar \pb)$, i.e., Algorithm \ref{alg: SCA}, quickly converges with around 15 iterations whereas that using $\hat R_s ( \xb_r, \bar \pb)$ takes around 100 iterations to reach the same value of sum rate. These numerical results corroborate Proposition \ref{prop: curvature}.
	
\vspace{-1em}	
\subsection{Double-Loop Accelerated Gradient Projection}  \label{sec:mulUE:surro3} 
	As discussed in Remark \ref{rmk1 inner subp}, for the convex approximation problem \eqref{mulUE_opt:2:app}, one may solve its Lagrange dual problem via the DSA method.
	However, due to the iterative SCA updates in Algorithm \ref{alg: SCA}, the DSA algorithm needs to be called for every iteration of SCA. Recently, the authors of  \cite{Shao:2018uj} proposed an algorithm that combines the SCA approximation  and the FISTA-type accelerated gradient projection (AGP) algorithm \cite{BeckFISTA2009}.
	The algorithm, which is referred to as gradient extrapolated majorization-minimization (GEMM), can in practice converge faster than the algorithm that uses AGP to solve the convex approximation problem in every SCA iteration. 
	In this subsection, we extend the idea of GEMM to our UAV JPPC problem \eqref{mulUE_opt:noPu}, and propose a double-loop AGP method.
	
	For ease of exposition, let us write problem \eqref{mulUE_opt:noPu} compactly as
	\begin{align}
	\max_{\yb}~  R_s(\yb) ~~{\rm s.t.~} \yb\in  \Yc,
	\end{align}
	where $\yb\triangleq (\xb_r^T, \left( \pb_r^{\rm U}\right) ^T, \left( \pb_r^{\rm D}\right) ^T, \pb_b^T)^T$, and
	\begin{align}\label{Y set}
	\Yc \triangleq \bigg\{
	\yb~ |~ &  \oneb^T \pb_{r}^{U} +\oneb^T  \pb_{r}^{\rm D} + \frac{\gamma_c}{\xi}\|\xb_r - \bb\|^2 \le P_r, 
	\notag \\
	&\left.    \oneb^T \pb_{b}+ \frac{\gamma_c}{\xi}\|\xb_r - \bb\|^2 \le P_b ,~ \pb\geq 0    \right\}.
	\end{align} Moreover, we write the surrogate function $\bar R_s(\xb_r,\pb)$ in \eqref{mulUE_opt:2:app:obj} as $\bar R_s(\yb; \bar \yb)$.
	By \cite{Shao:2018uj} , the GEMM involves the following iterative updates: for $i=1,2,\ldots,$
	\begin{align} \label{updt_gemm 1}
	\zb^{i}&=\yb^{i} +\frac{i-1}{i+2} \left( \yb^{i} - \yb^{i-1}\right),\\ \label{updt_gemm 2}
	\yb^{i+1}&= \Pi_{\Yc}\left[  \zb^i + \frac{1}{\tau_i}\nabla_{\yb}\bar {R}_s\left( \zb^i; \yb^i\right)  \right],
	\end{align}
	where $\tau_i>0$ is a step size which satisfies 
	\begin{align} \label{cvg_condition}
	\bar {R}_s\left( \yb^{i+1}; \yb^{i}\right) &\ge  \bar {R}_s\left( \zb^{i} ; \yb^i \right)  + 
	\nabla_{\yb}\bar {R}_s^T\left( \zb^i; \yb^i \right) \left(  \yb^{i+1} -\zb^i \right)
	- \frac{\tau_i}{2} \|\yb^{i+1} -\zb^i \|^2,
	\end{align}
	$\Pi_{\Yc}$ is the projection operation onto the set of $\Yc$, and it is defined that
	\begin{align}
	\zb^i \triangleq (( \zb_x^i) ^T, ( \zb_p^i) ^T)^T = ( (\zb_x^i)^T, ( \zb_p^{{\rm U},i}) ^T,( \zb_p^{{\rm D},i}) ^T,( \zb_p^{{\rm b},i}) ^T) ^T,\\
	\nabla_{\yb}\bar {R}_s\left( \zb^i; \yb^i \right) \triangleq {
		\left[ \begin{array}{ccc}
		\nabla_{\xb_r}\bar {R}_s\left( \zb^i; \yb^i \right)\\
		\nabla_{\pb}\bar {R}_s\left( \zb^i; \yb^i \right) 
		\end{array} 
		\right ]} = {
		\left[ \begin{array}{ccc}
		\nabla_{\xb_r}\bar {R}_s\left( \zb^i; \yb^i \right)\\
		\nabla_{\pb_r^{\rm U}}\bar {R}_s\left( \zb^i; \yb^i \right)\\
		\nabla_{\pb_r^{\rm D}}\bar {R}_s\left( \zb^i; \yb^i \right)\\
		\nabla_{\pb_b}\bar {R}_s\left( \zb^i; \yb^i \right)
		\end{array} 
		\right ]}.
	\end{align}
	As seen, the GEMM algorithm would be computationally efficient if \eqref{updt_gemm 2} admits a closed-form solution, e.g., when the constraint set $\Yc$ is simple such as box constraints.
	
	While the set $\Yc$ in \eqref{Y set} is not simple, \eqref{updt_gemm 2} may still be handled efficiently since it is a convex quadratically constrained quadratic program (QCQP). In particular, let us write  \eqref{updt_gemm 2}  explicitly as
	\begin{subequations}\label{updt_gemm 1 projection}  
		\begin{align}\label{updt_gemm 1 projection1}  
		\yb^{i+1} =\arg \min_{\xb_r, \pb \geq 0} ~&\|\xb_r - ( \zb_x^i + \frac{1}{\tau_i}\nabla_{\xb_r}\bar {R}_s\left( \zb^i; \yb^i\right) )  \|^2
		+\|\pb - ( \zb_p^i + \frac{1}{\tau_i}\nabla_{\pb}\bar {R}_s\left( \zb^i; \yb^i\right) )  \|^2 \\ \label{}
		\text{s.t.}~ & \mathbf{1}^T \pb_r^{\rm U} + \mathbf{1}^T \pb_r^{\rm D} + \frac{\gamma_c}{\xi}\|\xb_r - \bb\|^2 \le P_r, 
		\label{updt_gemm 1 projection2}   \\ \label{}
		& \oneb^T \pb_{b} + \frac{\gamma_c}{\xi}\|\xb_r - \bb\|^2 \le P_b. \label{updt_gemm 1 projection3}  
		\end{align}
	\end{subequations}
	Let $\lambda\geq 0$ and $\mu\geq 0$ be the dual variables associated with \eqref{updt_gemm 1 projection1}  and \eqref{updt_gemm 1 projection2}, respectively, and let $\nu_{rk}^{\rm D},\nu_{rk}^{\rm U},\nu_{bk}\ge 0$ to be the dual variables associated with constraints $p_{rk}^{\rm D},p_{rk}^{\rm U},p_{bk}\ge 0$ for all $k \in \Kc$. The Lagrangian of (\ref{updt_gemm 1 projection}) is 
	
	\begin{equation} \label{}
	\begin{split}
	\mathcal{L}^i\left(\yb, \lambda, \mu, \nub\right)  & = \|\xb_r - ( \zb_x^i + \frac{1}{\tau_i}\nabla_{\xb_r}\bar {R}_s\left( \zb^i; \yb^i\right) ) \|^2 + \|\pb - ( \zb_p^i + \frac{1}{\tau_i}\nabla_{\pb}\bar {R}_s\left( \zb^i; \yb^i\right) )\|^2 \\
	& + \lambda \left( \mathbf{1}^T \pb_r^{\rm U} + \mathbf{1}^T \pb_r^{\rm D} + \frac{\gamma_c}{\xi}\|\xb_r - \bb\|^2 - P_r\right) \\
	& + \mu \left(\mathbf{1}^T \pb_b + \frac{\gamma_c}{\xi}\|\xb_r - \bb\|^2 - P_b \right) - (\nub_r^{\rm D})^T \pb_r^{\rm D} - (\nub_r^{\rm U})^T \pb_r^{\rm U} - (\nub_b)^T \pb_b.
	\end{split}
	\end{equation}
	Then, the dual problem of (\ref{updt_gemm 1 projection}) is given by
	\begin{equation} \label{updt_gemm dual_problem}
	\max_{(\lambda,\mu, \nub) \ge 0} ~\left\lbrace \min_{\substack{\yb } } \mathcal{L}^i\left(\yb, \lambda, \mu, \nub\right) \right\rbrace \triangleq \max_{\substack{ (\lambda,\mu, \nub) \ge 0 }} g^i(\lambda, \mu, \nub),
	\end{equation}
	where $g(\lambda, \mu, \nub)$ is the dual function and can be obtained as
	\begin{equation} \label{updt_gemm dualfunc}
	\begin{split}
	g^i(\lambda, \mu, \nub) = & \frac{(\lambda + \mu)\frac{\gamma_c}{\xi}}{1 + (\lambda + \mu)\frac{\gamma_c}{\xi}} \|\bb - ( \zb_x^i + \frac{1}{\tau_i}\nabla_{\xb_r}\bar {R}_s\left( \zb^i; \yb^i\right) )\|^2 \\
	& - \frac{1}{4}\|\lambda \mathbf{1} - \nub_r^{\rm U}\|^2 + ( \zb_p^{{\rm U},i} + \frac{1}{\tau_i}\nabla_{\pb_r^{\rm U}}\bar {R}_s\left( \zb^i; \yb^i\right) )^T (\lambda \mathbf{1} - \nub_r^{\rm U})\\
	& - \frac{1}{4}\|\lambda \mathbf{1} - \nub_r^{\rm D}\|^2 + ( \zb_p^{{\rm D},i} + \frac{1}{\tau_i}\nabla_{\pb_r^{\rm D}}\bar {R}_s\left( \zb^i; \yb^i\right) )^T (\lambda \mathbf{1} - \nub_r^{\rm D})\\
	& - \frac{1}{4}\|\mu \mathbf{1} - \nub_b\|^2 + ( \zb_p^{b,i} + \frac{1}{\tau_i}\nabla_{\pb_b}\bar {R}_s\left( \zb^i; \yb^i\right) )^T (\mu \mathbf{1} - \nub_b) \\
	& - \lambda P_r - \mu P_b.
	\end{split}
	\end{equation}
	It is interesting to see that the dual function has a closed-form expression, thanks to the quadratic objective function and constraints in (\ref{updt_gemm 1 projection}). Instead of (\ref{updt_gemm 1 projection}), we solve the dual problem (\ref{updt_gemm dual_problem}). Specifically, since (\ref{updt_gemm dual_problem}) is a smooth convex optimization problem, we propose to solve \eqref{updt_gemm dual_problem} using the AGP method \cite{BeckFISTA2009}. Once (\ref{updt_gemm dual_problem}) is solved, the corresponding primal solutions to (\ref{updt_gemm 1 projection}) are the unique minimizer of $\min_{\yb} \Lc^i(\yb, \lambda, \mu, \nub)$, which are given by
	\begin{subequations} \label{updt_gemm privar}
		\begin{align}
		\xb_r & = \frac{ \zb_x^i + \frac{1}{\tau_i}\nabla_{\xb_r}\bar {R}_s\left( \zb^i; \yb^i\right)  + (\lambda + \mu)\frac{\gamma_c}{\xi}\bb}{1+ (\lambda + \mu)\frac{\gamma_c}{\xi}},\\
		{\pb_{r}^{\rm U}} & =  \zb_p^{{\rm U},i} + \frac{1}{\tau_i}\nabla_{\pb_r^{\rm U}}\bar {R}_s\left( \zb^i; \yb^i\right)  -\frac{1}{2} (\lambda \mathbf{1} - \nub_r^{\rm U}),\\
		{\pb_{r}^{\rm D}} & =  \zb_p^{{\rm D},i} + \frac{1}{\tau_i}\nabla_{\pb_r^{\rm D}}\bar {R}_s\left( \zb^i; \yb^i\right)-\frac{1}{2} (\lambda \mathbf{1} - \nub_r^{\rm D}),\\
		{\pb_b} & =  \zb_p^{b,i} + \frac{1}{\tau_i}\nabla_{\pb_b}\bar {R}_s\left( \zb^i; \yb^i\right)  -\frac{1}{2} (\mu \mathbf{1} - \nub_b).
		\end{align}
	\end{subequations}
	In summary, by combining the GEMM algorithm in \eqref{updt_gemm 1} and \eqref{updt_gemm 2}, and using the dual AGP algorithm to handle \eqref{updt_gemm 2},  we obtain Algorithm \ref{alg:double_loop} for solving our UAV JPPC problem (\ref{mulUE_opt:noPu}) which involves double loops of AGP steps. 
	In Algorithm \ref{alg:double_loop}, we denote $\sb = (\lambda, \mu, \nub^T)^T, g^i(\lambda, \mu, \nub) = g^i(\sb; \zb^i,\yb^i,\tau_i)$, and $\Sc = \left\lbrace \sb| \lambda \ge 0, \mu \ge 0, \nub \geq 0\right\rbrace $, for notation simplicity. 
	\begin{algorithm}[h] 
		\caption{Proposed Double-Loop AGP  for Problem (\ref{mulUE_opt:noPu}) }
		\begin{algorithmic}[1] \label{alg:double_loop}
			\STATE {\bf Initilize } $i = 1, \kappa, \tau_1$ and $\yb^1 = \yb^{0}$ that is feasible to (\ref{mulUE_opt:noPu});
			\REPEAT
			\STATE $\zb^i = \yb^i + \frac{i-1}{i+2}(\yb^i - \yb^{i-1})$;
			\STATE Initialize $\ell = 1, \sb^1 = \sb^{0} = \mathbf{0}$;
			\REPEAT
			\STATE $\tb^\ell = \sb^\ell + \frac{\ell-1}{\ell+2}(\sb^\ell - \sb^{\ell-1})$;
			\STATE $\sb^{\ell+1} = \Pi_{\Sc}\left[\tb^\ell + \frac{1}{\eta_l} \nabla_{\sb}g^i(\tb^\ell; \zb^i, \yb^i, \tau_i) \right] $, where $\eta_l$ is obtained by backtracking line search;
			\STATE $\ell \leftarrow \ell + 1$;
			\UNTIL $\frac{|g^i(\sb^\ell; \zb^i, \yb^i, \tau_i) - g^i(\sb^{\ell-1}; \zb^i, \yb^i, \tau_i)|}{|g^i(\sb^{\ell-1}; \zb^i, \yb^i, \tau_i)|}\leq \epsilon_1$.
			\STATE {Obtain} $\yb^{i+1}$ by (\ref{updt_gemm privar}) using $\sb^{\ell+1} = (\lambda^{\ell+1}, \mu^{\ell+1}, (\nu^{\ell+1})^T)^T$ and $(\zb^i, \yb^i)$;
			\IF{(\ref{cvg_condition}) is not met}
			\STATE $\tau_{i} \leftarrow \kappa \tau_i$, $\yb^{i}\leftarrow \yb^{i+1}$,  go to step 4;
			\ENDIF
			\STATE {$ \tau_{i+1} \leftarrow \tau_{i}$, $i \leftarrow i + 1$;}
			\UNTIL ${|R_s(\yb^{i})-R_s(\yb^{i-1})|}\leq \epsilon_2$. 
		\end{algorithmic}
	\end{algorithm}
	
	Before ending this section, we have the following remarks regarding two future directions.
	\begin{remark}
		\rm
		Different from the JPPC problem \eqref{mulUE_opt:1}, an alternative problem formulation is to consider minimizing the network sum power (the BS and the UAV) subject to individual rate constraint for each UE, i.e.,
		\begin{subequations} \label{mulUE_ qos}
			\begin{align} 
			\min_{\substack{ \xb_r, ~ \pb_b,\pb_r^{\rm U},\pb_r^{\rm D},\pb_u \geq  0,\\  p_{c} \ge 0} }~ & 
			\bigg[ \oneb^T \pb_{r}^{U} + \oneb^T\pb_{r}^{\rm D} + p_{c}\bigg] + \bigg[\oneb^T \pb_{b} + p_{c} \bigg]
			\\ \label{qos cons:ctrl 2v}
			\text{s.t.}~~ &  
			R_k^{\rm U}(\xb_r, p_{r,k}^{\rm U},p_{u,k}) + R_k^{\rm D}(\xb_r, p_{b,k},p_{r,k}^{\rm D})\geq R_{k,th},~\forall k\in \Kc, 
			\\
			& p_{u,k} \le  P_{u,k}, k \in \mathcal{K}, \\ 
			& {\rm SNR}_c \ge  \gamma_c,\label{qos cons:ctrl}
			\end{align} 
		\end{subequations}
		where $R_{k,th}$ is the minimum rate requirement of each UE $k$. 
		As seen, the proposed surrogate functions in Section \ref{sec:mulUE:surro1} can still be applied to \eqref{qos cons:ctrl 2v}, and the SCA algorithm \cite{SCA_1} can be used. However, the double-loop AGP algorithm is no longer applicable. 
		It is therefore interesting to investigate a computationally efficient algorithm to solve \eqref{mulUE_ qos} since \eqref{mulUE_ qos} has a large number of complex non-convex constraints. 
	\end{remark}
	
	\begin{remark}
		\rm Like the majority of the literature, the current work has assumed the LOS channels and fixed the flying altitude of the relay UAV. It is known that, under a more realistic probabilistic LOS channel model \cite{M.M_global15,A.A_WCL14}, the flying altitude directly affects the probability of LOS and non-LOS links. Here let us show the challenges for solving the JPPC problem if the probabilistic LOS channel model is considered. Denote $\beta_{\chi}$, $\chi\in \{\rm LOS,NLOS\}$, as the average path loss for LOS and NLOS channels \cite{M.M_global15}. Using the uplink link \eqref{rx:ul:relay}-\eqref{snr:ul:BS} as the example, the average uplink rate for UE $k$ is given by
		\begin{align}\label{average rate}
		\sum_{\chi, \chi' \in  \{\rm LOS,NLOS\} }P_{\chi}(\xb_r,\ub_k)P_{\chi'}(\xb_r,\bb)
		\log\bigg(1+ \frac{\beta_\chi \beta_{\chi'} p_{r,k}^{\rm U}  p_{u,k}  d_{kr}^{-2} d_{rb}^{-2}}{\beta_{\chi'} p_{r,k}^{\rm U} d_{rb}^{-2}\sigma^2 + \beta_\chi p_{u,k} d_{kr}^{-2}\sigma^2 + (\sigma^2)^2}\bigg)
		\end{align}
		where $P_{\rm NLOS}(\xb_r,\ub_k) = 1- P_{\rm LOS}(\xb_r,\ub_k) $, and according to  \cite{M.M_global15}.
		\begin{align}
		P_{\rm LOS}(\xb_r,\ub_k) =\frac{1}{1+a\exp\bigg(-b\bigg(\tan^{-1}\bigg(\frac{h}{\sqrt{(x_r-x_k)^2+(y_r-y_k)^2}}\bigg)-a\bigg)\bigg)},
		\end{align} where $a,b>0$ are some constants, is the probability to have LOS link between the UAV and UE $k$. The average rate in \eqref{average rate} includes the four combinations of the channel between the UAV and UE $k$ and that between the UAV and the BS.	
		While the proposed surrogate function in Section \ref{sec:mulUE:surro1} can be applied to approximate the log-term in \eqref{average rate}, the overall average rate would not be concave due to the additional probability functions. There needs new approximation techniques in order to handle the associated JPPC problem.
		
	\end{remark}
\section{Simulation Results \label{sec:simu}}
	In the section, simulation results are presented to evaluate the performance of the proposed algorithms. In the simulation the reference channel power gain at the distance $1~m$ from the transmitter is set to be $\beta = -40 ~\text{dB}$. The frequency bandwidth allocated to each UE is $W = 1~\text{MHz}$. The power spectrum density (PSD) of the noise power is $-169~\text{dBm/Hz}$. For convenience, tuples with the format $[x_{\text{min}},x_{\text{max}},y_{\text{min}}, y_{\text{max}}]$ are used to define the range of locations of the UEs and the BS. UEs are randomly located in the rectangular area $[0, 1000, 0, 1000]~m$, and the BS is randomly located in the rectangular area $[6000, 7000, 0, 1000]~m$. 
	The UEs and the BS are assumed to be on the ground. 
	The UAV hovers at a fixed altitude that is set as $h=100~m$.  Unless otherwise specified, the power budgets of each UE, the UAV and the BS are set to $P_{u,k} = 23 ~\text{dBm}, P_r = 36 ~\text{dBm}$ and  $ P_b = 43 ~\text{dBm}$, respectively.
	
\subsection{Convergence and Computation Time}	
	Similar to Fig. \ref{fig:surrogate_cvg_cmp}, we first examine the converge of the proposed SCA algorithm (Algorithm \ref{alg: SCA}) and double-loop AGP algorithm (Algorithm \ref{alg:double_loop}). For Algorithm \ref{alg: SCA}, we consider the use of \texttt{CVX} solver \cite{CVX} to solve problem \eqref{mulUE_opt:2:app} (denoted by Algorithm \ref{alg: SCA} + CVX), as well as the use of the DSA method described in Remark \ref{rmk1 inner subp} (denoted by Algorithm \ref{alg: SCA} + DSA). 
	The initial position of UAV $\xb_r$ is set to the geometry center $\mathbf{c} \triangleq (\frac{1}{K} \sum_{k=1}^K \ub_k + \bb)/2$, and the initial transmit powers of the UAV and that of the BS are uniformly allocated for each UE.
	The stopping criterion $\epsilon_0$ in Algorithm \ref{alg:mulULDL}  is set to $10^{-3}$.  
	The initial conditions for Algorithm \ref{alg:double_loop} are the same as those for Algorithm \ref{alg: SCA}, with additional 
	parameter $\kappa$ set to $1.2$, and $\tau_1$ is chosen such that the inequality
	\eqref{cvg_condition} holds for $\yb^1$ and $\yb^0$. 
	The stopping criterion in Algorithm \ref{alg:double_loop} is set to $\epsilon_1= 5\times 10^{-3}$ and $\epsilon_2= 10^{-3}$.
	
\begin{figure}[t]
		\centering 
		\vspace{0.2cm}	
		{\subfigure[][]
			{\resizebox{.49\textwidth}{!}
				{\includegraphics{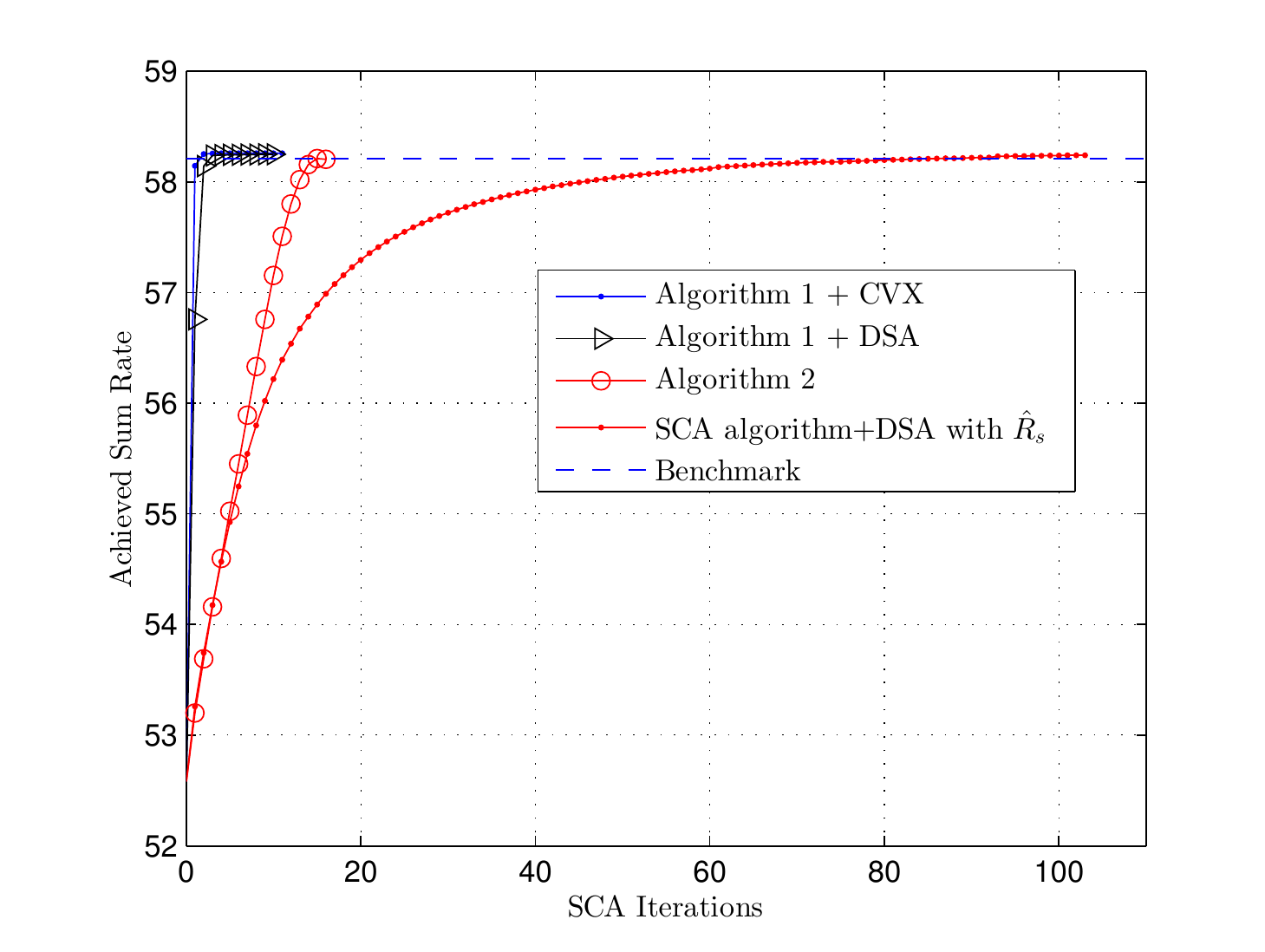}}} }			
		{\subfigure[][]
			{\resizebox{.49\textwidth}{!}
				{\includegraphics{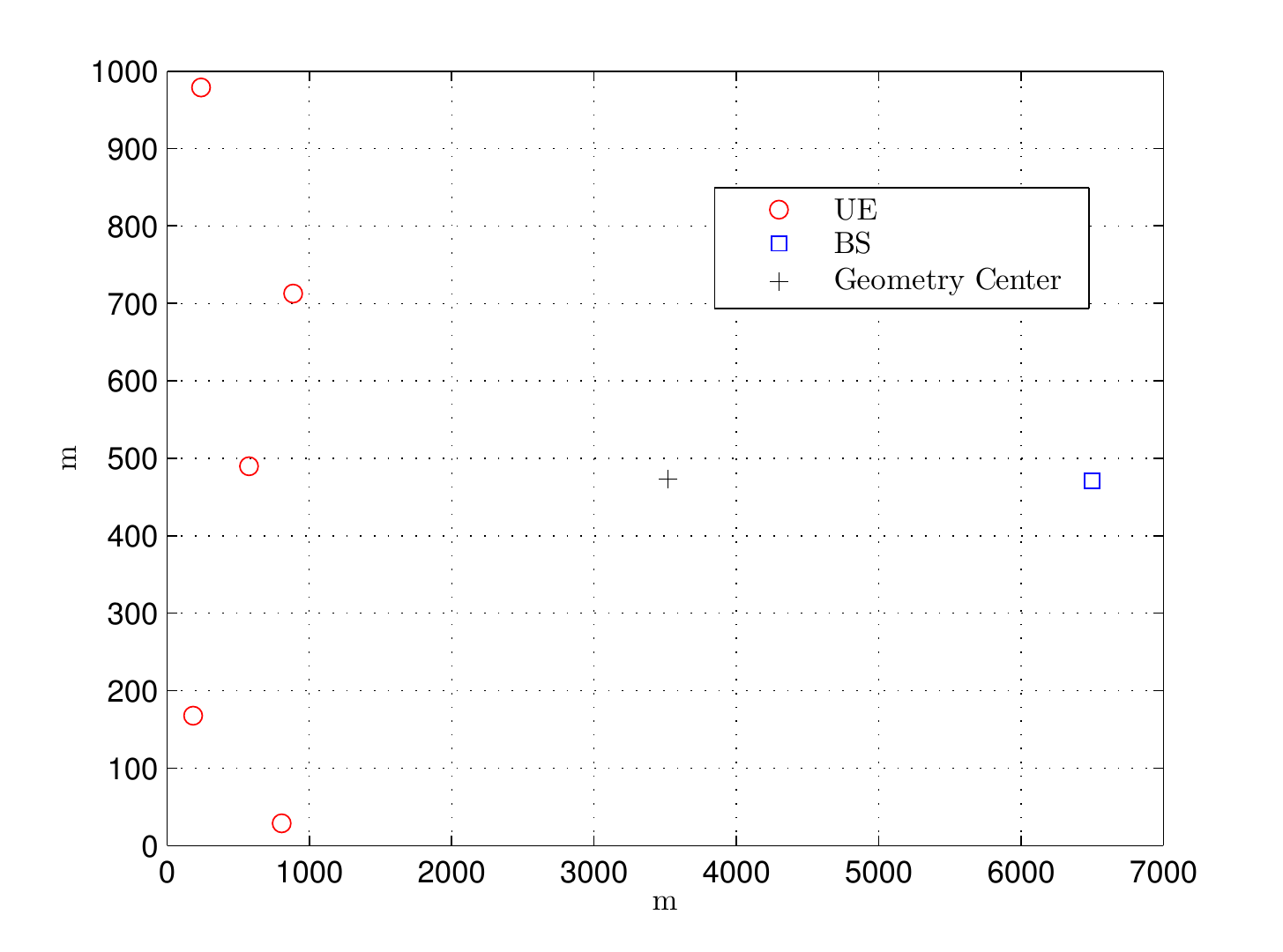}}} }
		\vspace{-0.5cm}
		\caption{(a) Sum rate versus iterations of SCA; $K=5$, $\gamma_c = 20~\text{dB}$.
			(b) Top view of the topology with $K = 5$ UEs.}
		\vspace{-0.3cm}	\label{fig:conv of sca}
	\end{figure}

\begin{figure}[t]
		\centering 
		\vspace{0.2cm}	
		{\subfigure[][]
			{\resizebox{.49\textwidth}{!}
				{\includegraphics{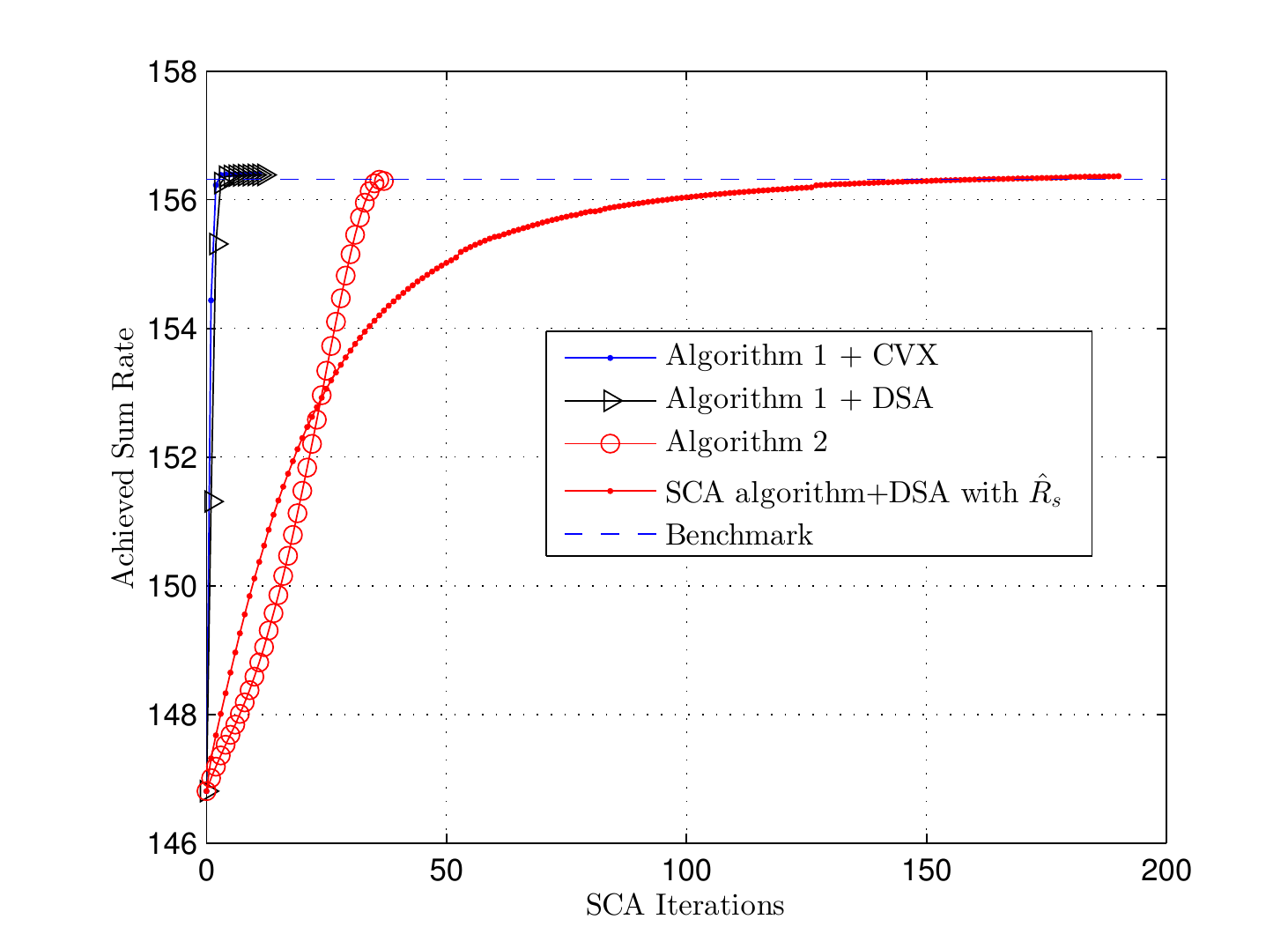}}} }			
		{\subfigure[][]
			{\resizebox{.49\textwidth}{!}
				{\includegraphics{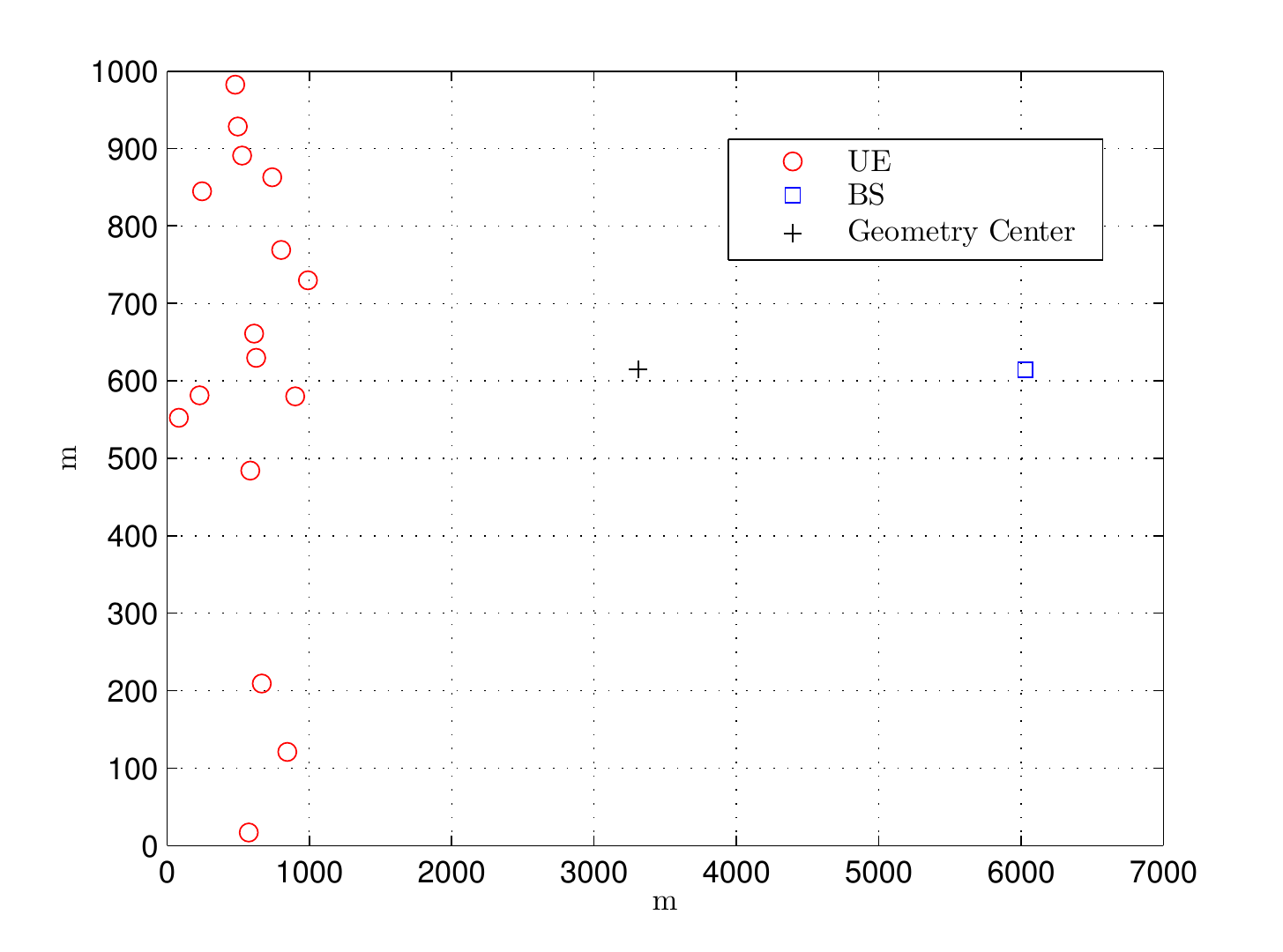}}} }
		\vspace{-0.5cm}
		\caption{(a) Sum rate versus number of SCA iterations; $K=16$, $\gamma_c = 20~\text{dB}$.
			(b) Top view of the topology with $K = 16$ UEs.}
		\vspace{-0.3cm}	\label{fig:conv of sca:nUE16}
	\end{figure}

	Fig. \ref{fig:conv of sca}(a) shows the achieved sum rates versus the SCA iteration of Algorithm \ref{alg: SCA} + {CVX}, Algorithm \ref{alg: SCA} + DSA, and Algorithm \ref{alg:double_loop}.
	The SCA algorithm using $\hat R_s$ as the surrogate function and using DSA for solving \eqref{mulUE_opt:2:app} is also presented, denoted by `SCA algorithm  +  DSA with $\hat R_s$'.	
	The curve of benchmark is the converged sum rate achieved by Algorithm \ref{alg: SCA} + {CVX}.
	The locations of UEs and BS are shown in Fig. \ref{fig:conv of sca}(b) for $K=5$ and $\gamma_c = 20~\text{dB}$.
	One can see that the sum rates increase with the iteration numbers.
	The curve of Algorithm \ref{alg: SCA} + DSA is similar to that of Algorithm \ref{alg: SCA} + {CVX}.
	They respectively take 10 iterations and 11 iterations to reach the benchmark.
	Interestingly, Algorithm \ref{alg:double_loop} can even converge faster than the SCA algorithm  +  DSA with $\hat R_s$. 
	Fig. \ref{fig:conv of sca:nUE16} displays similar results for a scenario with 16 UEs $K=16$.
	
	\begin{table}[!t]\centering
		\caption{ Comparison on the average computation time.}
		\begin{tabular}{lclcl} \toprule
			&Number of UEs ($K$) & 5  & 10  & 16  \\\midrule
			&SCA algorithm + DSA with $\hat R_s$   & 36.06 s  & 58.78 s    & 114.18 s\\
			&Algorithm \ref{alg:mulULDL} + DSA   & 4.28 s   & 7.21 s     & 11.38 s\\
			&Algorithm \ref{alg:double_loop} & 0.10 s   & 0.17 s     & 0.25 s\\
			\bottomrule
			\label{table:computeTime_cmp:new}
		\end{tabular}\vspace{-0.5cm}
	\end{table}
	In Table \ref{table:computeTime_cmp:new}, the average running time of SCA algorithm + DSA with $\hat R_s$, Algorithm \ref{alg:mulULDL} + DSA, and Algorithm \ref{alg:double_loop} are presented for different numbers of UEs. The results are obtained by averaging 70 random simulation trials, conducted on a laptop computer with a 2-core 2.50 GHz CPU and 4 GB RAM.
	As seen, consistent with the results in Fig. \ref{fig:conv of sca}(a) and Fig. \ref{fig:conv of sca:nUE16}(a), Algorithm \ref{alg:mulULDL} + DSA is much more computationally efficient than the SCA algorithm + DSA with $\hat R_s$.
	Moreover, Algorithm \ref{alg:double_loop} is about 40 times faster than Algorithm \ref{alg:mulULDL} + DSA in terms of the running time.
	
\subsection{Performance of Wireless Two-Way Relaying}
\begin{figure}[t]
		\centering 
		\vspace{0.2cm}	
		{\subfigure[][]
			{\resizebox{.5\textwidth}{!}
				{\includegraphics{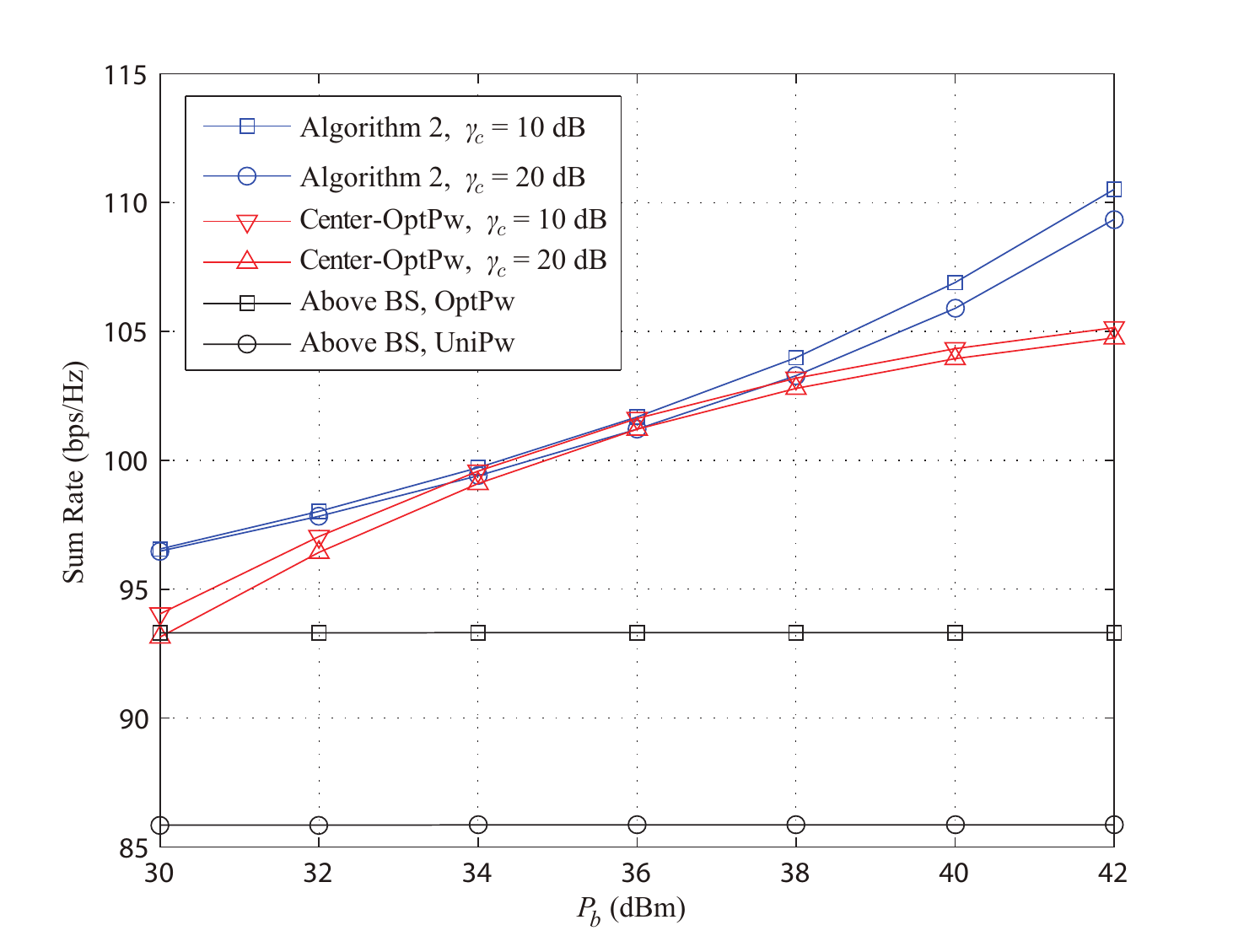}}} }			
		{\subfigure[][]
			{\resizebox{.48\textwidth}{!}
				{\includegraphics{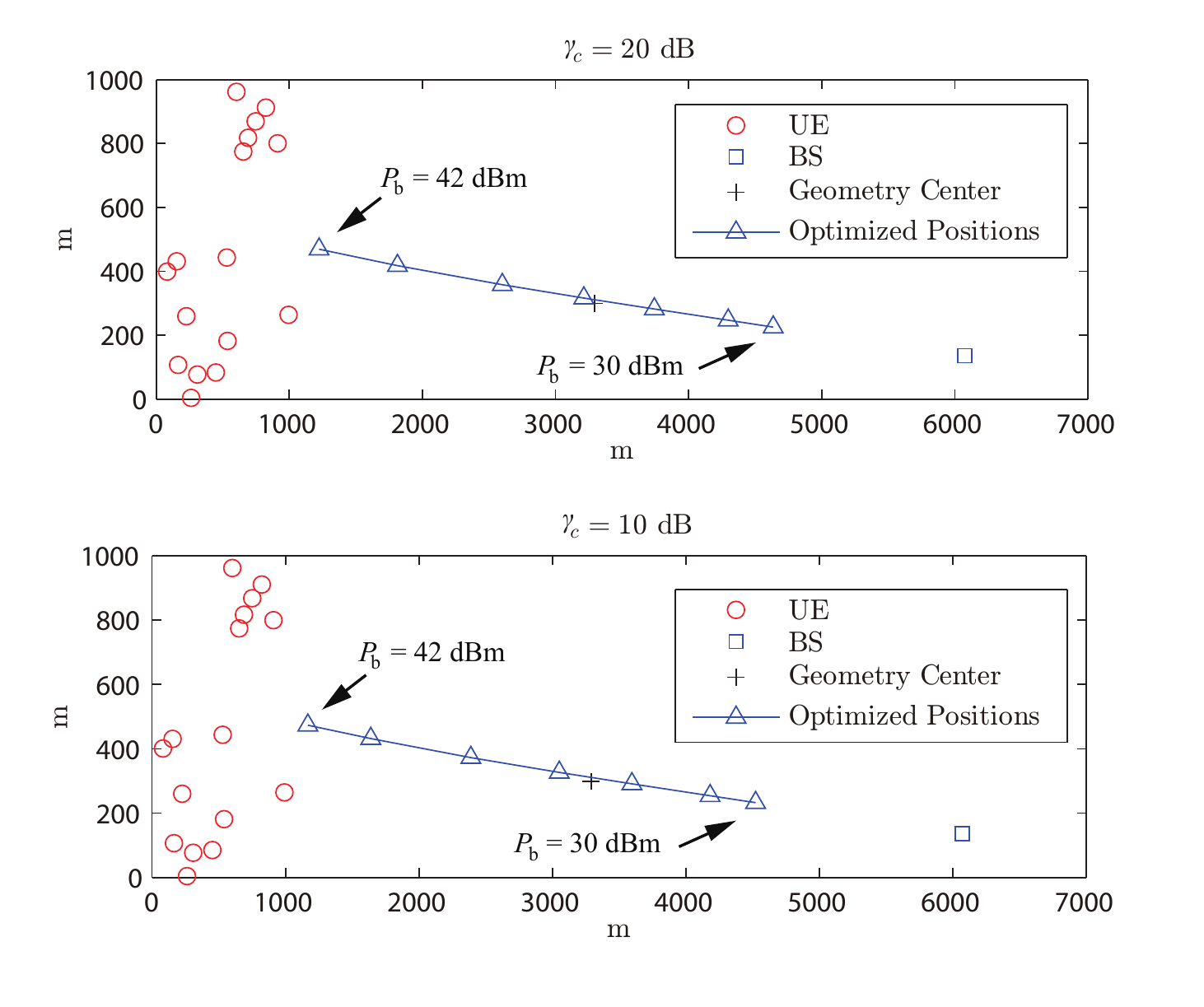}}} }
		\vspace{-0.5cm}
		\caption{(a) Sum rate versus the power budget of BS $P_b$. 
			(b) Top view of the topology with $K = 16$ UEs and the optimized UAV position under different BS power budget $P_b$}
		\vspace{-0.6cm}	\label{fig: performance Pb}
	\end{figure}

	{\bf Example 1:} In Fig. \ref{fig: performance Pb}, we display the achieved sum rate as well as the optimized UAV positions versus the BS transmit power budget $P_b$, for a scenario with 16 UEs $(K = 16)$ and control link constraint $\gamma_c=10$ and $20$ dB, respectively.
	Except for the proposed Algorithm \ref{alg:double_loop} which jointly optimizes the UAV position and transmission powers of all terminals, we also present the results that the UAV is fixed either on the top of the BS (Above BS) or at the geometry center (GeoCenter). When the UAV's position is fixed above BS, we either consider fixed uniform power allocation (UniPw) or optimized power allocation (OptPw) for the BS and UAV.

	One can observe in Fig. \ref{fig: performance Pb}(a) that the sum rates achieved by most of the methods significantly increase with a larger $P_b$, whereas increments of the sum rates by `Above BS, OptPw' and `Above BS, UniPw' are negligible.
	It can also be seen that the proposed Algorithm \ref{alg:double_loop} can always achieve the best performance, comparing to all the other methods since in Algorithm \ref{alg:double_loop} the transmission power and the UAV position are jointly optimized.
	In Fig. \ref{fig: performance Pb}(b), it is observed that the optimized position of UAV tends to move closer to UEs if a larger $P_b$ is given. 
	As the optimized UAV positions are around the geometry center when $P_b$ ranges from 34 dB to 38 dB, the sum rate achieved by Algorithm \ref{alg:double_loop} is only slightly higher than that achieved by the method `Center-OptPw' as seen from Fig. \ref{fig: performance Pb}(a).
	
	To examine the impact of the control link, the optimal positions achieved by Algorithm \ref{alg:mulULDL} under different values of $\gamma_{c}$ are also given in Fig. \ref{fig: performance Pb}(b). One can see that with a less stringent control link SNR requirement ($\gamma_c = 10~\text{dB}$), the UAV can move further from the BS and closer to UEs, bringing a higher sum rate as shown in Fig. \ref{fig: performance Pb}(a).  Thus, the control link plays an important role in the resource allocation of the UAV-enabled relaying communications and should not be overlooked.
	   \begin{figure}[h]
		\centering 
		\vspace{0.2cm}	
		{\subfigure[][]
			{\resizebox{.5\textwidth}{!}
				{\includegraphics{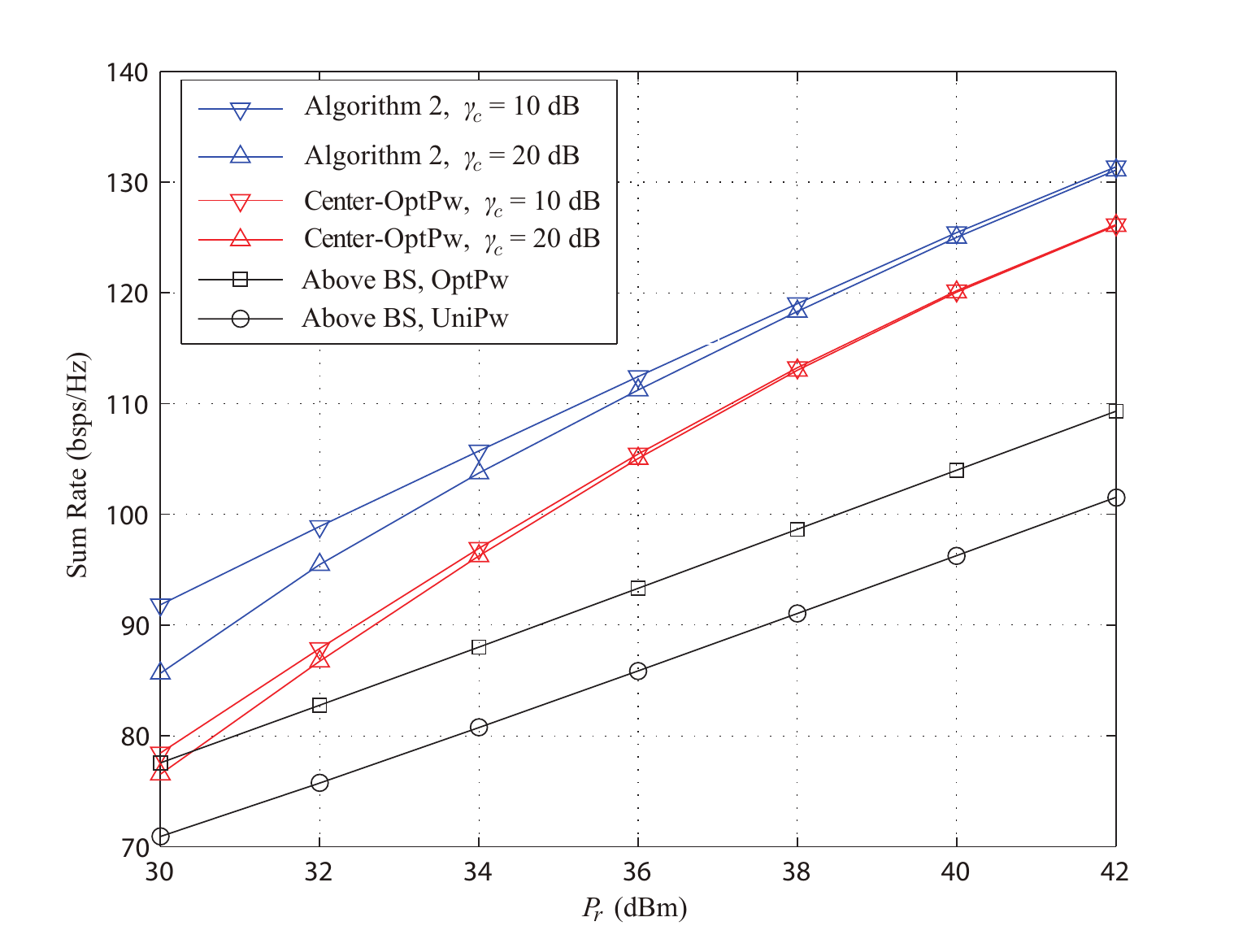}}} }			
		{\subfigure[][]
			{\resizebox{.48\textwidth}{!}
				{\includegraphics{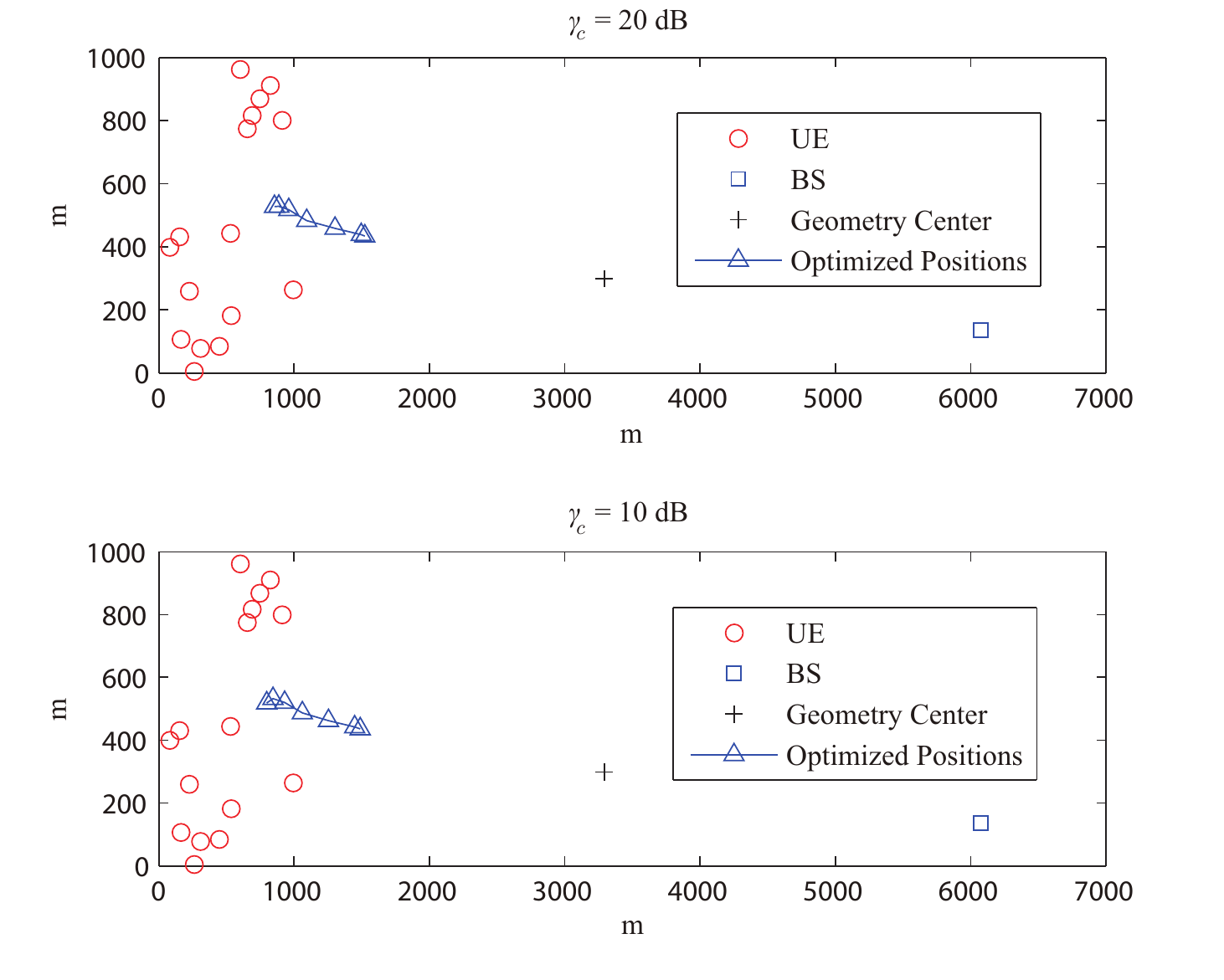}}} }
		{\subfigure[][]
			{\resizebox{.5\textwidth}{!}
				{\includegraphics{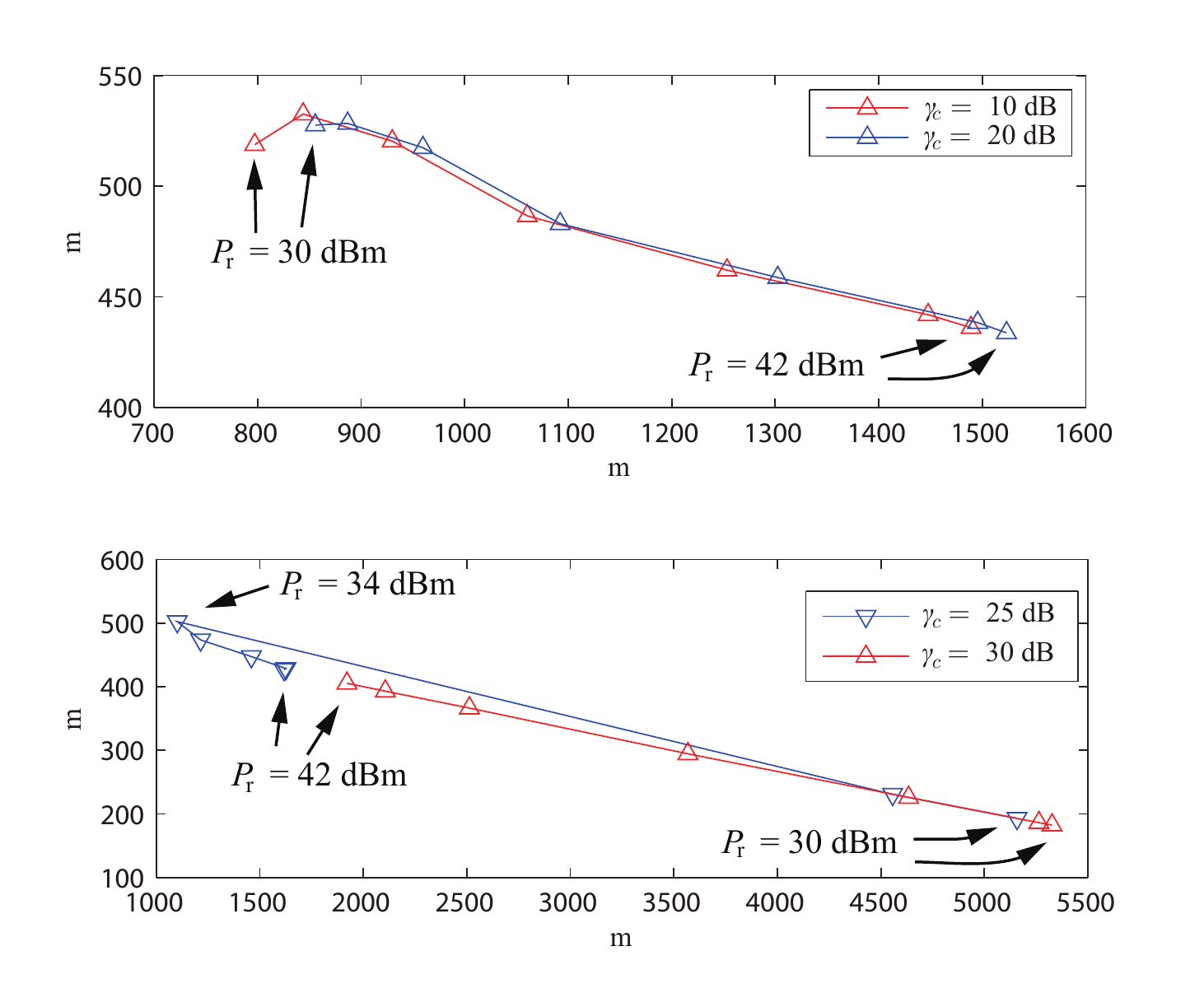}}} }
		\vspace{-0.1cm}
		\caption{(a) Sum rate versus the power budget of UAV $P_r$. 
			(b), (c) Top view of the topology with $K = 16$ UEs and the optimized UAV position under different UAV power budget $P_r$ and control link SNR requirement $\gamma_c$.}
		\vspace{-0.7cm}	\label{fig: performance Pr}
	\end{figure}

	{\bf Example 2:} 
	In Fig. \ref{fig: performance Pr}(a), the sum rates versus the UAV budget $P_r$ achieved by various schemes under consideration are displayed. 
	The topology of UEs and BS are shown in Fig. \ref{fig: performance Pr}(b). One can see that the sum rates obtained by all the algorithms increase with increasing $P_r$ since more power can be allocated for information relaying. Besides, the proposed Algorithm \ref{alg:double_loop} is superior than the other two schemes with fixed UAV position. One can also observe from the figure that the sum rate achieved by Algorithm \ref{alg:double_loop} under $\gamma=10$ dB is higher than that under $\gamma=20$ dB, especially when $P_r$ is smaller. This is because the percentage of power that needs to be allocated for the control link is much larger when $P_r$ is smaller and $\gamma=20$ dB.
	
		In Figs. \ref{fig: performance Pr}(c), the optimized UAV positions under different values of $P_r$ and $\gamma_c$ are presented. 
	One can observe from Fig. \ref{fig: performance Pr}(c) that when $\gamma_{c} = 10~\text{dB}$ or $\gamma_{c} = 20~\text{dB}$, the optimized position of UAV will move closer to UE when $P_r$ increases from $30~\text{dBm}$ to $42~\text{dBm}$. However, as seen from Figures \ref{fig: performance Pr}(c), under $\gamma_c = 30~\text{dB}$, the moving direction of the optimized UAV position is opposite -- it moves toward to the BS when $P_r$ increases. 
	However, given $\gamma_c = 25~\text{dB}$, the optimized UAV position first moves closer to the UE, and then turns back to the BS at $P_r = 34 ~\text{dBm}$. 
	In other words, the optimized UAV position is not trivially monotonic with respect to $P_r$.
	The reason is that the transmit power at the UAV is not only related to the control link signaling, but also affects the uplink and downlink data transmission.

\section{Conclusion \label{sec:conclusion}}
	In the paper, we have investigated the JPPC problem \eqref{mulUE_opt:1} to maximize the sum rate of the UAV-enabled wireless two-way relay network. While the formulated problem \eqref{mulUE_opt:1} has a complicated sum rate function and is not concave, we have proposed the concave surrogate function in Proposition \ref{prop surrogate func}, and shown theoretically and numerically that the concave surrogate function can provide a significantly faster SCA convergence.
	To further improve the computational efficiency, we have exploited the quadratic constraint structure of \eqref{mulUE_opt:noPu} and developed a double-loop AGP algorithm (Algorithm \ref{alg:double_loop}). The double-loop AGP algorithm has a computation time that is at least an order of magnitude less than its counterpart based on SCA and DSA method. Moreover, the presented simulation results have shown how the BS power budget,  UAV power budge and SNR requirement on the control link can affect the optimal relay UAV positioning. In particular, it is shown that the optimal UAV positioning is non-trivial since the optimized network sum rate can be greatly higher than those obtained by simple positioning strategies.

\appendices
\vspace{-0.3cm}
\section{Proof of Property \ref{lemma: x_r}} \label{proof of property 2}
	Denote $\ssb=\frac{\bb - \ub}{\|\bb - \ub\|} \triangleq (s_x, s_y, 0)^T,$ and let $\ssb_{\perp}\triangleq (s'_x, s'_y, 0)^T$ be orthogonal to $\ssb$ and satisfy $\|\ssb_{\perp}\|=1$.
	Then, the position of the UAV can be expressed as $\xb_r=\ub + \alpha \ssb + \alpha_{\perp} \ssb_{\perp} + h \eb_z$, where $\eb_z=(0,0,1)^T$ and $\alpha,\alpha_{\perp} \in \mathbb{R}$.
	For the denominator of the first logarithmic term in \eqref{sigUE_opt:uldl:obj1}, we can bound it as follows
	\begin{align}
	&p_{r}^{\rm U} d_{ur}^2 +  P_{u} d_{rb}^2+ \xi^{-1}d_{ur}^{2} d_{rb}^{2}    \notag \\
	=  &~ p_{r}^{\rm U} \| \xb_r -\ub \|^2 + P_u \| \xb_r - \bb \|^2 + \xi^{-1} \| \xb_r -\ub \|^2 \| \xb_r - \bb \|^2 
	\notag  
	\\
	=  &~ p_{r}^{\rm U} \| \alpha \ssb + \alpha_{\perp} \ssb_{\perp} + h \eb_z \|^2  \notag  \\
	& ~~+ P_u \| \ub + \alpha \ssb + \alpha_{\perp} \ssb_{\perp} + h \eb_z- \bb \|^2 \notag  \\ 
	&~~ +\xi^{-1} \| \alpha \ssb + \alpha_{\perp} \ssb_{\perp} + h \eb_z \|^2~ \| \ub + \alpha \ssb + \alpha_{\perp} \ssb_{\perp} + h \eb_z - \bb \|^2  
	\notag 
	\\
	= & ~p_{r}^{\rm U} (\alpha^2 + \alpha_{\perp}^2 + h^2)+ P_u ((M-\alpha)^2 + \alpha_{\perp}^2 + h^2)  \notag  \\
	&~~ + \xi^{-1}(\alpha^2 + \alpha_{\perp}^2 + h^2)((M-\alpha)^2 + \alpha_{\perp}^2 + h^2)  \notag  \\
	\geq & ~ p_{r}^{\rm U} (\alpha^2 + h^2)+ P_u ((M-\alpha)^2 + h^2)  +\xi^{-1} (\alpha^2 + h^2)((M-\alpha)^2 + h^2),
	\label{lower bound}
	\end{align}
	where $M=\|\bb-\ub\|$, and the last inequality is obtained by setting $ \alpha_{\perp}=0$.
	In addition, the denominator of the second logarithmic term in \eqref{sigUE_opt:uldl:obj1}	can have a similar lower bound.
	Therefore, the sum rate in \eqref{sigUE_opt:uldl:obj1} would be maximized if $\xb_r=\ub + \alpha \ssb + h \eb_z$.
	Besides, for $\alpha \geq M$, the optimal $\alpha$ that minimizes the lower bound in \eqref{lower bound} is $\alpha=M$.
	So, the optimal UAV position, when projected onto the x-y plane, must be on the line segment connecting the UE and the BS. 
	\hfill $\blacksquare$

\section{ \label{proof of surrogate functions} Proof of Proposition \ref{prop surrogate func}}
	Note that 
	%
	\begin{align}
	\label{lbd:R_kD:1}
	R_k^{\rm D} \left( p_{b,k}, p_{r,k}^{\rm D}, \xb_r\right) & = \text{log} \left(1 + \frac{ \xi^2 p_{r,k}^{\rm D}  p_{b,k} d_{rb}^{-2} d_{kr}^{-2}}{ \xi p_{r,k}^{\rm D} d_{kr}^{-2} +  \xi p_{b,k} d_{rb}^{-2} + 1} \right) 
	\\ 
	&= \text{log} \left( 1+ \frac{1}{\xi^{-1} \frac{\|\db_{rb}\|^2}{p_{b,k}} + \xi^{-1} \frac{\|\db_{kr}\|^2}{p_{r,k}^{\rm D}} + \xi^{-2} \frac{\|\db_{rb}\|^2}{p_{b,k}}\frac{\|\db_{kr}\|^2}{p_{r,k}^{\rm D}} } \right)  \notag 
	\\ 
	& = \text{log}\left( 1 + \xi^{-1} \frac{\|\db_{rb}\|^2}{p_{b,k}}\right) + \text{log}\left( 1 +  \xi^{-1} \frac{\|\db_{kr}\|^2}{p_{r,k}^{\rm D}}\right) 
	\notag 
	\\
	&~~~~~~~~~~~~~~  -\text{log}\left( \xi^{-1} \frac{\|\db_{rb}\|^2}{p_{b,k}} + \xi^{-1} \frac{\|\db_{kr}\|^2}{p_{r,k}^{\rm D}} + \xi^{-2} \frac{\|\db_{rb}\|^2}{p_{b,k}}\frac{\|\db_{kr}\|^2}{p_{r,k}^{\rm D}} \right). 
	\label{lbd:R_kD:99}
	\end{align}
	Since $\frac{\|{\mathbf{x}}\|^2}{p}$ is convex in $\mathbf{x} \in \mathbb{R}^n$ and $p>0$, its first-order approximation at a given point $(\bar{\mathbf{x}},\bar p)$ is a lower bound, i.e.,
	\begin{equation}\label{eqn: lower bound x2overy}
	\frac{\|{\mathbf{x}}\|^2}{p} \ge \frac{2 \bar{\mathbf{x}}^T}{\bar p} \mathbf{x} - \frac{\|\bar{\mathbf{x}}\|^2}{{\bar{p}}^2}p.
	\end{equation}
	Thus, the first and second terms in the right hand side (RHS) of \eqref{lbd:R_kD:99} can be bounded as
	\begin{align}
	\text{log}\left( 1 + \xi^{-1} \frac{\|\db_{rb}\|^2}{p_{b,k}}\right) &\geq  \text{log}\left( 1 + \xi^{-1} \left(   \frac{2\bar\db_{rb}^T}{\bar p_{b,k}}\db_{rb} - \frac{\|\bar\db_{rb}\|^2}{\bar p_{b,k}^2}  p_{b,k}\right) \right), \label{bound1} \\
	\text{log}\left( 1 +  \xi^{-1} \frac{\|\db_{kr}\|^2}{p_{r,k}^{\rm D}}\right) &\geq  \text{log}\left( 1 + \xi^{-1} \left(  \frac{2\bar\db_{kr}^T}{\bar p_{r,k}^{\rm D}}\db_{kr}  - \frac{\|\bar\db_{kr}\|^2}{(\bar p_{r,k}^{\rm D})^2} p_{r,k}^{\rm D}\right) \right), \label{bound2}
	\end{align}
	where both of the lower bounds are concave functions.
	Since $-\text{log}(1+x)$ is convex satisfying $-\text{log}(x) \ge -\text{log}(\bar{x}) + \frac{\bar{x}-x}{\bar{x}}$	for any $\bar x >0$, the third term in the RHS of \eqref{lbd:R_kD:99} can be bounded as
	\begin{align}
	&-\text{log}\left( \xi^{-1} \frac{\|\db_{rb}\|^2}{p_{b,k}} + \xi^{-1} \frac{\|\db_{kr}\|^2}{p_{r,k}^{\rm D}} + \xi^{-2} \frac{\|\db_{rb}\|^2}{p_{b,k}}\frac{\|\db_{kr}\|^2}{p_{r,k}^{\rm D}} \right)  \notag \\
	&\geq - \text{log}(\bar I_k^{\rm D}) + 1 -\frac{1}{\bar I_k^{\rm D}}\left( \xi^{-1} \frac{\|\db_{rb}\|^2}{p_{b,k}} + \xi^{-1} \frac{\|\db_{kr}\|^2}{p_{r,k}^{\rm D}} + \xi^{-2} \frac{\|\db_{rb}\|^2}{p_{b,k}}\frac{\|\db_{kr}\|^2}{p_{r,k}^{\rm D}} \right),\label{bound3}
	\end{align}
	where $\bar I_k^{\rm D}$ is given in \eqref{IkD}. 
	
	By applying 
	\begin{equation}
	xy = \frac{1}{2}\left( x+y\right)^2 - \left( \frac{1}{2} x^2 + \frac{1}{2} y^2\right) ,
	\end{equation}
	to the term $-\frac{\|\db_{rb}\|^2}{p_{b,k}} \frac{\|\db_{kr}\|^2}{p_{r,k}^{\rm D}}$ in the RHS of \eqref{bound3}, we have
	\begin{align}
	-\frac{\|\db_{rb}\|^2}{p_{b,k}} \frac{\|\db_{kr}\|^2}{p_{r,k}^{\rm D}} = 
	-\frac{1}{2}  \left(  \frac{\|\db_{rb}\|^2}{p_{b,k}} + \frac{\|\db_{kr}\|^2}{p_{r,k}^{\rm D}} \right) ^2 + \frac{1}{2}  \left(  \frac{\|\db_{rb}\|^2}{p_{b,k}}\right) ^2 
	+\frac{1}{2}  \left( \frac{\|\db_{kr}\|^2}{p_{r,k}^{\rm D}}\right) ^2.  \label{bound4}
	\end{align}
	In addition, by applying the first-order condition of the convex function $\frac{\left( \|\mathbf{x}\|^2\right)^2}{p^2} $, i.e., 
	\begin{equation}
	\frac{\left( \|\mathbf{x}\|^2\right)^2}{p^2} \ge \frac{4\|\bar{\mathbf{x}}\|^2 \bar{\mathbf{x}}^T \mathbf{x}}{\bar p^2}- \frac{2 (\|\bar{\mathbf{x}}\|^2)^2 p}{\bar p^3}  - \frac{(\|\bar{\mathbf{x}}\|^2)^2}{\bar p^2},
	\end{equation} to the last two terms in the RHS of \eqref{bound4}, we can further obtain a lower bound of \eqref{bound3} as
	\begin{align}
	&-\text{log}\left( \xi^{-1} \frac{\|\db_{rb}\|^2}{p_{b,k}} + \xi^{-1} \frac{\|\db_{kr}\|^2}{p_{r,k}^{\rm D}} + \xi^{-2} \frac{\|\db_{rb}\|^2}{p_{b,k}}\frac{\|\db_{kr}\|^2}{p_{r,k}^{\rm D}} \right)  \notag \\
	&\geq 
	- \text{log}(\bar I_k^{\rm D}) + 1 - \frac{1}{\bar I_k^{\rm D} \xi} \left( \frac{\|\db_{rb}\|^2}{p_{b,k}} +  \frac{\|\db_{kr}\|^2}{p_{r,k}^{\rm D}} \right) - \frac{1}{2 \bar I_k^{\rm D} \xi^2}  \left(  \frac{\|\db_{rb}\|^2}{p_{b,k}} + \frac{\|\db_{kr}\|^2}{p_{r,k}^{\rm D}} \right) ^2 \notag 
	\\  
	&~~~ + \frac{1}{2 \bar I_k^{\rm D} \xi^2} \bigg(  \left( \frac{4\|\bar\db_{rb}\|^2 \bar\db_{rb}^T \db_{rb}}{\bar p_{b,k}^2}- \frac{2 (\|\bar\db_{rb}\|^2)^2 p_{b,k}}{\bar p_{b,k}^3}  - \frac{(\|\bar\db_{rb}\|^2)^2}{\bar p_{b,k}^2} \right) 
	\notag \\
	&~~~~~~~~~~~~~~~~~~~~+\left(   \frac{4\|\bar\db_{kr}\|^2 \bar\db_{kr}^T \db_{kr}}{(\bar p_{r,k}^{\rm D})^2}- \frac{2 (\|\bar\db_{kr}\|^2)^2 p_{r,k}^{\rm D}}{(\bar p_{r,k}^{\rm D})^3}  - \frac{(\|\bar\db_{kr}\|^2)^2}{(\bar p_{r,k}^{\rm D})^2}\right)   \bigg).
	\label{bound5}
	\end{align}
	By substituting \eqref{bound1}, \eqref{bound2} and \eqref{bound5} into \eqref{lbd:R_kD:99}, one then obtains \eqref{surrogate func D}.
	It is easy to check that 
	\begin{align}
	&R_k^{\rm D} ( \bar\xb_r, \bar p_{b,k}, \bar p_{r,k}^{\rm D}) = \bar R_k^{\rm D} ( \bar\xb_r, \bar p_{b,k}, \bar p_{r,k}^{\rm D}).
	\label{surrogate func D tight}
	\end{align}	
	Equation \eqref{RkUs} can be derived in an analogous fashion; the details are skipped here.
	%
	\hfill $\blacksquare$

\section{\label{app:cmp_surrogate12}Derivation of \eqref{surrogate D2} and \eqref{surrogate U2}}	
	Let us introduce auxiliary variables  $a_{r,k}^{\rm D} = \sqrt{p_{r,k}^{\rm D}}, a_{b,k} = \sqrt{p_{b,k}}, s_{kr} = d_{kr}^2$, $s_{rb} = d_{rb}^2$, and $J_k^{\rm D} = \xi (\frac{(a_{r,k}^{\rm D})^2}{s_{kr}} + \frac{a_{b,k}^2}{ s_{rb}})$. 
	Moreover, for a given feasible point $(\bar \pb_b,\bar \pb_r^{\rm D},\bar \xb_r)$, we define $\bar a_{r,k}^{\rm D}=\sqrt{\bar p_{r,k}^{\rm D}}, \bar a_{b,k}=\sqrt{\bar p_{b,k}} $, $\bar s_{kr} = \bar d_{kr}^2$, $\bar s_{rb} = \bar d_{rb}^2$, and $\bar J_k^{\rm D} = \xi (\frac{(\bar a_{r,k}^{\rm D})^2}{\bar s_{kr}} + \frac{\bar a_{b,k}^2}{\bar  s_{rb}})$.
	We can bound $R_k^{\rm D} \left( p_{b,k}, p_{r,k}^{\rm D}, \xb_r\right)=R_k^{\rm D} \left( a_{b,k}, a_{r,k}^{\rm D}, \xb_r\right)$ as follows
	\begin{align} 
	\!\!\!R_k^{\rm D} \left( a_{b,k}, a_{r,k}^{\rm D}, \xb_r\right)  &= \text{log} \left( 1 + \frac{ \xi^2 \frac{ a_{b,k}^2 (a_{r,k}^{\rm D})^2}{s_{rb} s_{kr}} }{ \xi \frac{(a_{r,k}^{\rm D})^2}{s_{kr}} + \xi \frac{a_{b,k}^2}{s_{rb}} + 1 } \right) \notag \\
	&= \text{log} \left( 1 + \xi \frac{(a_{r,k}^{\rm D})^2}{s_{kr}} + \xi \frac{a_{b,k}^2}{s_{rb}} + \xi^2 \frac{ a_{b,k}^2 (a_{r,k}^{\rm D})^2}{s_{rb} s_{kr}} \right) - \text{log} \left(1 + \xi \frac{(a_{r,k}^{\rm D})^2}{s_{kr}} + \xi \frac{a_{b,k}^2}{s_{rb}}   \right) \notag \\
	&= \text{log} \left( 1 + \xi \frac{(a_{r,k}^{\rm D})^2}{s_{kr}} \right) + \text{log} \left( 1 + \xi \frac{a_{b,k}^2}{s_{rb}} \right)- \text{log} \left(1 + \xi \frac{(a_{r,k}^{\rm D})^2}{s_{kr}} + \xi \frac{a_{b,k}^2}{s_{rb}}   \right) \label{old_lbd:R_kD0} \\
	& \ge \text{log} \left( 1 + \xi \frac{2 \bar a_{r,k}^{\rm D} }{\bar s_{kr} } a_{r,k}^{\rm D} - \xi \frac{ (\bar{a}_{r,k}^{\rm D})^2 }{\bar s_{kr}^2 } s_{kr} \right) + \text{log} \left( 1 + \xi \frac{2 \bar a_{b,k} }{\bar s_{rb} } a_{b,k} - \xi \frac{ (\bar a_{b,k})^2 }{\bar s_{rb}^2 } s_{rb} \right) \notag  \\
	& ~~~ - \text{log} \left( 1 + \bar{J}_k^{\rm D}\right) + \frac{\bar{J}_k^{\rm D}}{1 + \bar{J}_k^{\rm D}} - \frac{\xi}{1 + \bar{J}_k^{\rm D}}\bigg(\frac{(a_{r,k}^{\rm D})^2}{\|{\mathbf{x}}_r - {\ub}_k \|^2} + \frac{a_{b,k}^2}{ \|{\mathbf{x}}_r - {\bb} \|^2 }\bigg) \label{old_lbd:R_kD}
	\end{align}
	\begin{align}
	& \ge \text{log} \left( 1 + \xi \frac{2 \bar a_{r,k}^{\rm D} }{\bar s_{kr} } a_{r,k}^{\rm D} - \xi \frac{ (\bar{a}_{r,k}^{\rm D})^2 }{\bar s_{kr}^2 } \|{\mathbf{x}}_r - {\ub}_k \|^2 \right) \notag  \\
	& ~~~+\text{log} \left( 1 + \xi \frac{2 \bar a_{b,k} }{\bar s_{rb} } a_{b,k} - \xi \frac{ (\bar a_{b,k})^2 }{\bar s_{rb}^2 } \|{\mathbf{x}}_r - {\bb} \|^2 \right)
	- \text{log} \left( 1 + \bar{J}_k^{\rm D}\right) + \frac{\bar{J}_k^{\rm D}}{1 + \bar{J}_k^{\rm D}} \notag \\
	&~~~ - \frac{\xi}{1 + \bar{J}_k^{\rm D}} \left( \frac{(a_{r,k}^{\rm D})^2}{\| {\ub}_k\|^2- \|\bar {\mathbf{x}}_r\|^2 + 2 (\bar {\mathbf{x}}_r - {\ub}_k)^T {\mathbf{x}}_r }  + \frac{a_{b,k}^2}{\| {\bb}\|^2 - \|\bar {\mathbf{x}}_r\|^2 + 2 (\bar {\mathbf{x}}_r - {\bb})^T {\mathbf{x}}_r} \right), \label{old_lbd:R_kD1}
	\end{align}
	which is the surrogate function in \eqref{surrogate D2}.
	To obtain \eqref{old_lbd:R_kD}, we have applied \eqref{eqn: lower bound x2overy} to $\frac{(a_{r,k}^{\rm D})^2}{s_{kr}}$ and $\frac{a_{b,k}^2}{s_{rb}}$ in the first two logarithmic terms in \eqref{old_lbd:R_kD0}; we have also applied the first-order Taylor lower bound of 
	$-\text{log}(1+x) \geq -\text{log}(1+\bar x) - \frac{x -\bar x}{1+\bar x}$ to the third logarithmic term in \eqref{old_lbd:R_kD0}.
	The inequality in \eqref{old_lbd:R_kD1} is obtained by applying the first-order Taylor lower bound to $\|{\mathbf{x}}_r - {\ub}_k \|^2$ and $\|{\mathbf{x}}_r - {\bb} \|^2$, respectively. 
	
	%
	
	The surrogate function in \eqref{surrogate U2} for $\hat R_k^{\rm U} ( \xb_r, a_{r,k}^{\rm U})$ can be obtained in a similar fashion and the details are omitted here.
	\hfill $\blacksquare$

\section{Proof of Proposition \ref{prop: curvature}} \label{proof:Hessian_cmp}
	Since $\bar{R}_k^{\rm D}$ (resp. $\hat {R}_k^{\rm D}$) and $\bar{R}_k^{\rm U}$ (resp. $\hat {R}_k^{\rm U}$) have the same structure, we only consider $\bar{R}_k^{\rm D}$ and $\hat{R}_k^{\rm D}$ in the proof.
	Note that $\bar{R}_k^{\rm D}$ and $\hat{R}_k^{\rm D}$ are concave functions, and thereby their Hessian matrices are negative semi-definite, i.e., $-\nabla_{\mathbf{x}_r}^2\bar{R}_k^{\rm D} \left( \xb_r,\pb  \right)\succeq \zerob$ and 
	$-\nabla_{\mathbf{x}_r}^2\hat{R}_k^{\rm D} \left( \xb_r,\pb  \right)\succeq \zerob$.
	
	By \eqref{RkDs}, the negative Hessian of $\bar{R}_k^{\rm D}(\xb_r,\bar \pb)$ with respect to $\xb_r$ can be derived as
	%
	\begin{align}
	\!\!\!\!\!\!&-\nabla_{\mathbf{x}_r}^2\bar{R}_k^{\rm D} \left( \xb_r,\bar \pb  \right) \notag \\
	& = \frac{4 \bar{\db}_{rb} \bar{\db}_{rb}^T}{\left( \xi \bar p_{b,k} + 2 \bar{\db}_{rb}^T \db_{rb} -\|\bar{\db}_{rb}\|^2 \bar p_{b,k} \right) ^2}  
	+ \frac{4 \bar{\db}_{kr} \bar{\db}_{kr}^T}{\left( \xi \bar p_{r,k}^{\rm D} + 2 \bar{\db}_{kr}^T \db_{kr} -\|\bar{\db}_{kr}\|^2 \bar p_{r,k}^{\rm D} \right) ^2} \notag \\
	& ~~~ + \frac{2}{\bar{I}_k^{\rm D} \xi}  \left( \frac{1}{\bar p_{b,k}} + \frac{1}{\bar p_{r,k}^{\rm D}}\right) \Ib \notag \\
	&
	~~~ + \frac{2}{\bar{I}_k^{\rm D} \xi^2}  \bigg( \frac{ 2 \db_{rb}\db_{rb}^T + \|\db_{rb}\|^2\Ib}{\bar p_{b,k}^2}   
	+ \frac{  2 \db_{rb}\db_{kr}^T + \|\db_{rb}\|^2 \Ib}{\bar p_{b,k}\bar  p_{r,k}^{\rm D}}  +\frac{2 \db_{kr}\db_{rb}^T + \|\db_{kr}\|^2 \Ib}{\bar p_{r,k}^{\rm D} \bar p_{b,k}} +\frac{  2 \db_{kr}\db_{kr}^T + \|\db_{kr}\|^2 \Ib}{(\bar p_{r,k}^{\rm D})^2}\bigg),
	\label{hessian of bar R}
	\end{align}
	where $\Ib$ is the identity matrix and $\bar{I}_k^{\rm D} $ is given in \eqref{IkD} which is a function of $\xi$.
	As seen, when $\xi$ is large, the first two terms in the right hand side (RHS) of \eqref{hessian of bar R} is bounded by $\mathcal{O}(\xi^{-2})\Ib$, the last four terms in the RHS of \eqref{hessian of bar R} is bounded by $\mathcal{O}(\xi^{-1})\Ib$, and 
	the third term in the RHS of \eqref{hessian of bar R} is bounded by 
	\begin{align}
	\frac{2}{\bar{I}_k^{\rm D} \xi}  \left( \frac{1}{\bar p_{b,k}} + \frac{1}{\bar p_{r,k}^{\rm D}}\right) \Ib	  
	&\preceq
	\frac{2}{ \frac{\|\bar{ \db}_{rb}\|^2}{\bar p_{b,k}} + \frac{\|\bar{ \db}_{kr}\|^2}{\bar p_{r,k}^D} } \left( \frac{1}{\bar p_{b,k}} + \frac{1}{\bar p_{r,k}^{\rm D}}\right) \Ib \notag \\
	& = 2\bigg(\frac{ \frac{\|\bar{ \db}_{rb}\|^2}{\bar p_{b,k}} + \frac{\|\bar{ \db}_{kr}\|^2}{\bar p_{r,k}^D} }{ \frac{1}{\bar p_{b,k}} + \frac{1}{\bar p_{r,k}^{\rm D}}} \bigg)^{-1}	\Ib 
	\preceq  \frac{2}{\min\left\lbrace \|\bar{ \db}_{rb}\|^2, \|\bar{ \db}_{kr}\|^2\right\rbrace } \Ib .
	\end{align}
	Thus, the maximum eigenvalue of $-\nabla_{\mathbf{x}_r}^2\bar{R}_k^{\rm D} \left( \xb_r,\bar \pb  \right)$ is bounded as
	\begin{align}\label{bound of bar R}
	\lambda_{\max}(-\nabla^2_{\xb_r} \bar R_k^{\rm D} ( \xb_r, \bar \pb)) \leq \mathcal{O}(\xi^{-1}) + \frac{2}{\min\left\lbrace \|\bar{ \db}_{rb}\|^2, \|\bar{ \db}_{kr}\|^2\right\rbrace }.
	\end{align}
	
	By \eqref{surrogate D2},  the negative Hessian matrix of $\hat{R}_k^{\rm D}\left( \xb_r,\bar \pb  \right) $ w.r.t. $\xb_r$ is
	\begin{align}
	-\nabla_{\mathbf{x}_{r}}^2\hat{R}_k^{\rm D}\left( \xb_r,\bar \pb  \right) & = \frac{4}{ \left( \frac{\|\bar{\db}_{rb}\|^4}{\xi \bar p_{b,k}} + 2 \|\bar{\db}_{rb}\|^2  - \|{\db}_{rb} \|^2 \right)^2} \db_{rb} \db_{rb}^T  
	+\frac{2}{ \frac{\|\bar{\db}_{rb}\|^4}{\xi \bar p_{b,k}} + 2 \|\bar{\db}_{rb}\|^2  - \|{\db}_{rb} \|^2 } \Ib  \notag \\
	& ~~~ + \frac{\xi}{1 + \bar{J}_k^{\text{D}}} \frac{8p_{b,k} \bar {\db}_{rb}\bar {\db}_{rb}^T}{ \left(2 \bar {\db}_{rb}^T\db_{rb} -\|\bar {\db}_{rb}\|^2 \right) ^3} \notag 
	\end{align}
	\begin{align}
	&~~~ + \frac{4}{ \left( \frac{\|\bar{\db}_{kr}\|^4}{\xi \bar p_{r,k}^{\rm D}} + 2 \|\bar{\db}_{kr}\|^2  - \|{\db}_{kr} \|^2 \right)^2} \db_{kr} \db_{kr}^T  
	+\frac{2}{ \frac{\|\bar{\db}_{kr}\|^4}{\xi \bar p_{r,k}^{\rm D}} + 2 \|\bar{\db}_{kr}\|^2  - \|{\db}_{kr} \|^2 } \Ib  \notag \\
	& ~~~ + \frac{\xi}{1 + \bar{J}_k^{\text{D}}} \frac{8p_{r,k}^{\rm D} \bar {\db}_{kr}\bar {\db}_{kr}^T}{ \left(2 \bar {\db}_{kr}^T\db_{kr} -\|\bar {\db}_{kr}\|^2 \right) ^3},\label{hessian of hat R}
	\end{align}
	where $\bar J_k^{\rm D}$ is given in \eqref{bar J}.
	One can see that each term in the RHS of \eqref{hessian of hat R} is bounded by $\mathcal{O}(1)$ when $\xi$ is large.
	Thus, 
	\begin{align}
	\lambda_{\max}(-\nabla^2_{\xb_r} \hat R_k^{\rm D} ( \xb_r, \bar \pb)) \leq \mathcal{O}(1).
	\end{align}
	
	To show \eqref{comp of curvature}, let $\xb_r = \bar{\xb}_r$ in \eqref{hessian of hat R} and take $\xi\to \infty$. We then obtain
	\begin{align}\label{old_sf_ub}
	\lim_{\xi \to \infty}-\nabla_{\mathbf{x}_{r}}^2 \hat{R}_k^{\rm D}\left( \bar{\xb}_r,\bar{ \pb}  \right)   
	& = \frac{4}{  \|\bar{\db}_{rb}\|^6 } \bar{\db}_{rb} \bar{\db}_{rb}^T  
	+\frac{2}{  \|\bar{\db}_{rb}\|^2 } \Ib  + \frac{1}{\frac{\bar p_{r,k}^{\rm D}}{\|\bar{\db}_{kr}\|^2} + \frac{\bar p_{b,k}}{\|\bar{\db}_{rb}\|^2}} \frac{ 8 \bar p_{b,k} }{ \|\bar {\db}_{rb}\|^6}\bar {\db}_{rb}\bar {\db}_{rb}^T \notag \\
	& ~~~+ \frac{4}{ \|\bar{\db}_{kr}\|^4} \bar{\db}_{kr} \bar{\db}_{kr}^T  
	+\frac{2}{  \|\bar{\db}_{kr}\|^2 } \Ib + \frac{1}{\frac{\bar p_{r,k}^{\rm D}}{\|\bar{\db}_{kr}\|^2} + \frac{\bar p_{b,k}}{\|\bar{\db}_{rb}\|^2}} \frac{ 8 \bar p_{r,k}^{\rm D} }{ \|\bar {\db}_{kr}\|^6}\bar {\db}_{kr}\bar {\db}_{kr}^T \notag \\
	& \succ \frac{2}{  \|\bar{\db}_{rb}\|^2 } \Ib +\frac{2}{  \|\bar{\db}_{kr}\|^2 } \Ib \notag \\
	& \succ \frac{2}{ \min\left\lbrace { \|\bar{\db}_{rb}\|^2, \|\bar{\db}_{kr}\|^2 } \right\rbrace } \Ib. 
	\end{align}
	From the above lower bound and \eqref{bound of bar R}, we conclude that, when $\xi \to \infty$,
	\begin{align}
	\lambda_{\max}(-\nabla^2_{\xb_r} \hat R_k^{\rm D} ( \bar \xb_r, \bar \pb)) 
	\geq 
	\lambda_{\min}(-\nabla^2_{\xb_r} \hat R_k^{\rm D} ( \bar  \xb_r, \bar \pb)) 
	> \lambda_{\max}(-\nabla^2_{\xb_r} \bar R_k^{\rm D} ( \bar \xb_r, \bar \pb)).
	\end{align}
	The proof is complete. 	\hfill $\blacksquare$

\ifCLASSOPTIONcaptionsoff
  \newpage
\fi

\bibliography{Ref1}
\end{document}